\documentclass{aa} %[referee]
\usepackage[varg]{txfonts}
\usepackage{soul}
\usepackage{ifthen}
\usepackage{graphicx}
\usepackage{bbm}
\usepackage{Paper_Linc}
\usepackage[AA]{Math_Linc}
\defcitealias{Liu_etal_2014}{Liu, X. et al.}
\defcitealias{Liu_etal_2015a}{Liu, X. et al.}
\defcitealias{Liu_etal_2014a}{Liu, J. et al.}
\defcitealias{Liu_etal_2015}{Liu, J. et al.}
\defcitealias{Lin_Kilbinger_2015}{Paper I}
\defcitealias{Lin_Kilbinger_2015a}{Paper II}
\newcommand{\PaperI}{\citetalias{Lin_Kilbinger_2015}}
\newcommand{\PaperII}{\citetalias{Lin_Kilbinger_2015a}}
\newcommand{\MRLens}{\textsc{MRLens}}

\title{A new model to predict weak-lensing peak counts}
\subtitle{III. Filtering technique comparisons}
\titlerunning{A new model to predict weak-lensing peak counts III.}
\author{Chieh-An Lin\inst{\ref{inst1}}, Martin Kilbinger\inst{\ref{inst1},\ref{inst2}}, \and Sandrine Pires\inst{\ref{inst1}}}
\authorrunning{C.-A. Lin, M. Kilbinger, \& S. Pires}
\institute{
	Service d'Astrophysique, CEA Saclay, Orme des Merisiers, B\^at 709, 91191 Gif-sur-Yvette, France\label{inst1}\\
\email{\texttt{chieh-an.lin@cea.fr}}
	\and 
	Sorbonne Universités, UPMC Univ. Paris 6 et CNRS, UMR 7095, Institut
d'Astrophysique de Paris, 98 bis bd Arago 75014 Paris, France\label{inst2}
} 
\date{Received 21 March 2016 / Accepted 08 June 2016}

\abstract
	{This is the third in a series of papers that develop a new and flexible model to predict weak-lensing (WL) peak counts, which have been shown to be a very valuable non-Gaussian probe of cosmology.}
	{In this paper, we compare the cosmological information extracted from WL peak counts using different filtering techniques of the galaxy shear data, including linear filtering with a Gaussian and two compensated filters (the starlet wavelet and the aperture mass), and the nonlinear filtering method \MRLens. We present improvements to our model that account for realistic survey conditions, which are masks, shear-to-convergence transformations, and non-constant noise.}
	{We create simulated peak counts from our stochastic model, from which we obtain constraints on the matter density $\OmegaM$, the power spectrum normalisation $\sigEig$, and the dark-energy parameter $\wZero$. We use two methods for parameter inference, a copula likelihood, and approximate Bayesian computation (ABC). We measure the contour width in the $\OmegaM$-$\sigEig$ degeneracy direction and the figure of merit to compare parameter constraints from different filtering techniques.}
	{We find that starlet filtering outperforms the Gaussian kernel, and that including peak counts from different smoothing scales helps to lift parameter degeneracies. Peak counts from different smoothing scales with a compensated filter show very little cross-correlation, and adding information from different scales can therefore strongly enhance the available information. Measuring peak counts separately from different scales yields tighter constraints than using a combined peak histogram from a single map that includes multiscale information.}
	{Our results suggest that a compensated filter function with counts included separately from different smoothing scales yields the tightest constraints on cosmological parameters from WL peaks.}
	
\keywords{Gravitational lensing: weak, Cosmology: large-scale structure of Universe, Methods: statistical}

\begin{document}
\maketitle

\section{Introduction}
\label{sect:intro}

Without the need to assume any relationship between baryons and dark matter, weak gravitational lensing (WL) is directly sensitive to the total matter distribution. WL probes massive structures in the Universe on large scales, providing information about the late-time evolution of the matter, which helps analyzing the equation of state of dark energy. 

Recently, CFHTLenS \citep[][etc.]{Heymans_etal_2012, VanWaerbeke_etal_2013, Kilbinger_etal_2013, Erben_etal_2013, Fu_etal_2014} has shown that the third generation lensing surveys provide interesting results on cosmological constraints. While other surveys such as KiDS \citep{Kuijken_etal_2015}, DES \citep{TheDarkEnergySurveyCollaboration_etal_2015}, and the Subaru Hyper-Suprime Cam (HSC) survey are expected to deliver results in coming years, cosmologists also look forward to reaching higher precision with more ambitious projects like Euclid, LSST, and WFIRST.

Several methods to extract information from WL exist. Until now, a great focus has been put on two-point statistics, for example the matter power spectrum. This is motivated by the fact that the matter spectrum can be well modeled by theory on large scales. However, due to complex gravitational interactions, the matter distribution becomes nonlinear and non-Gaussian on small scales. In this case, not only the theoretical spectrum needs to be corrected \citep{Makino_etal_1992, Bernardeau_etal_2002, Baumann_etal_2012, Carrasco_etal_2012}, but also the rich non-Gaussian information is discarded. For these reasons, including non-Gaussian observables complementary to the power spectrum strongly enhances weak lensing studies.

A suitable candidate for extracting non-Gaussian information is WL peak counts. These local maxima of projected mass density trace massive regions in the Universe, and are thus a probe of the halo mass function. According to \citetalias{Liu_etal_2015} \citeyearpar{Liu_etal_2015}, it turns out that peak counts alone constrain cosmology better than the power spectrum, implying the importance of non-Gaussian observables. This strengthens the motivation for peak-count studies.

Previous analyses on peaks can be divided into two categories. The first category is concerned with \emph{cluster-oriented purposes}. Motivated to search for galaxy clusters using WL, these studies (e.g. \citealt{White_etal_2002}; \citealt{Hamana_etal_2004, Hamana_etal_2012}, \citeyear{Hamana_etal_2015}; \citealt{Hennawi_Spergel_2005}; \citealt{Schirmer_etal_2007}; \citealt{Gavazzi_Soucail_2007}; \citealt{Abate_etal_2009}) focus on very high peaks, in general with signal-to-noise ratio (S/N) larger than four, and study purity, completeness, positional offsets, the mass-concentration relation, etc. A cross-check with galaxy clusters is often done. On the other hand, the second category, which concerns \emph{cosmology-oriented purposes}, focuses on peaks with a wider range of S/N ($\gtrsim 1$). Peaks from this range can not necessarily be identified with massive clusters. They can also arise from large-scale structure projections, be spurious signals, or a mixture of all of these cases. Studies for this purpose model true and spurious peaks together and constrain cosmology. This second purpose is the focus of this paper.

For cosmology-oriented purposes, correctly predicting the total peak counts is essential. Until now, three methods have been proposed: analytical models (\citealt{Maturi_etal_2010, Maturi_etal_2011}; \citealt{Fan_etal_2010}; \citetalias{Liu_etal_2014} \citeyear{Liu_etal_2014}, \citeyear{Liu_etal_2015a}), modeling using $N$-body simulations (\citealt{Wang_etal_2009}; \citealt{Marian_etal_2009, Marian_etal_2010, Marian_etal_2011}, \citeyear{Marian_etal_2012}, \citeyear{Marian_etal_2013}; \citealt{Dietrich_Hartlap_2010}; \citealt{Kratochvil_etal_2010, Yang_etal_2011}, \citeyear{Yang_etal_2013}; \citealt{Bard_etal_2013}; \citetalias{Liu_etal_2014a} \citeyear{Liu_etal_2014a}, \citeyear{Liu_etal_2015}; \citealt{Martinet_etal_2015}), and fast stochastic forward modeling \citep{Lin_Kilbinger_2015, Lin_Kilbinger_2015a}. While analytical models struggle when confronted by observational effects, $N$-body simulations are very costly for parameter constraints. Motivated by these drawbacks, \citet[][hereafter \PaperI]{Lin_Kilbinger_2015} proposed a new model to predict WL peak counts, which is both fast and flexible. It has been shown that the new model agrees well with $N$-body simulations.

In WL, the convergence, which is interpreted as the projected mass, is not directly observable, while the (reduced) shear is. To reconstruct the mass, a common way is to invert the relation between convergence and shear \citep{Kaiser_Squires_1993, Seitz_Schneider_1995}. Then, to reduce the shape noise level, inverted maps are usually smoothed with a Gaussian kernel. However, inversion techniques create artefacts and modify the noise spectrum in realistic conditions. An alternative is to use the aperture mass \citep{Kaiser_etal_1994}, which applies a linear filter directly on the shear field. This is equivalent to filter the convergence with a compensated kernel.

Besides, there also exists various nonlinear reconstruction techniques. For example, \citet{Bartelmann_etal_1996} proposed to minimize the error on shear and magnification together. Other techniques are sparsity-based methods such as \textsc{MRLens} \citep{Starck_etal_2006}, \textsc{FASTLens} \citep{Pires_etal_2009}, and \textsc{Glimpse} \citep{Leonard_etal_2014}. These approaches aim to map the projected mass through a minimization process.

Among these different filtering methods, some studies for optimal peak selection, such as \citet{Maturi_etal_2005} and \citet{Hennawi_Spergel_2005}, have been made. These methods are optimal in different senses. On the one hand, \citet{Maturi_etal_2005} modeled large-scale structures as noise with respect to clusters. Following this reasoning, given a halo density profile on a given scale, they obtained the ideal shape for the smoothing kernel. On the other hand, \citet{Hennawi_Spergel_2005} constructed a tomographic matched filter algorithm. Given a kernel shape, this algorithm was able to determine the most probable position and redshift of presumed clusters. Actually, these two studies display two different strategies for dealing with multiple scales. The \emph{separated strategy} (followed implicitly by \citealt{Maturi_etal_2005}, \citeyear{Maturi_etal_2010}; see also \citetalias{Liu_etal_2015} \citeyear{Liu_etal_2015}) applies a series of filters at different scales. Cosmological constraints are then derived by combining the peak abundance information obtained in each filtered WL map. In the \emph{combined strategy} \citep[followed e.g. by][]{Hennawi_Spergel_2005, Marian_etal_2012}, sometimes called \emph{mass mapping}, the significance from different scales are combined into a single filtered map from which we estimate peak abundance  and derive constraints.

Up to now, the question of optimal filtering for cosmology-oriented purposes remains unsolved. For cluster-oriented purposes, the comparison is usually based on purity and completeness \citep{Hennawi_Spergel_2005, Pires_etal_2012, Leonard_etal_2014}. However, for cosmology-oriented purposes, since we are interested in constraining cosmological parameters, we should focus on indicators like the Fisher matrix, the figure of merit (FoM), etc. So far, no study has compared filtering techniques with regard to these indicators. This will be the approach that we adopt here for comparison.

In this paper, we address the following questions:
\begin{itemize}\itemsep=0.5ex
	\item For a given kernel shape, with the separated strategy, what are the preferable characteristic scales?
	\item Which can extract more cosmological information, the compensated or non-compensated filters?
	\item Which can extract more cosmological information, the separated or combined strategy?
	\item How do nonlinear filters perform?
\end{itemize}
To obtain the constraints, we use two statistical techniques: the copula likelihood and approximate Bayesian computation (ABC). To perform the comparison, we use two indicators to measure the tightness of constraints. An example for this methodology has been shown by \citet[][hereafter \PaperII]{Lin_Kilbinger_2015a}, on the comparison between different definitions of data vector. 

Compared to \PaperI\ and \PaperII, this study improves the model to account for more realistic observational features. We apply a redshift distribution for source galaxies, include masks, construct the convergence $\kappa$ from the reduced shear instead of computing $\kappa$ directly, test different filters, determine the noise level locally, and include the equation of state of dark energy for the constraints.

The paper is structured as follows. We begin with theoretical basics in \sect{sect:basics}. Then, we introduce the different filters used in this study in \sect{sect:filtering}. In \sect{sect:methodology}, we describe the methodology adopted in this study. In \sect{sect:results}, we show our results both from the likelihood and ABC. And finally, a discussion is presented in \sect{sect:summary}.

\section{Theoretical basics}
\label{sect:basics}

\subsection{Mass function}
%\label{sect:basics_massFct}

The halo mass function indicates the population of dark matter halos, depending on mass $M$ and redshift $z$. This variation is usually characterized by the quantity $f(\sigma)$ varying with regard to the density contrast dispersion of the matter field $\sigma(z,M)$. Defining $n(z, \lessM)$ as the halo number density at $z$ with mass less than $M$, the function $f$ is defined as
\begin{align}
	f(\sigma) \equiv \frac{M}{\bar{\rho}_0}\frac{\rmd n(z,\lessM)}{\rmd\ln\sigma\inv(z,M)},
\end{align}
where $\bar{\rho}_0$ is the background matter density at the current time. The quantity $\sigma(z,M)\equiv D(z)\sigma(M)$ can be furthermore defined as the product of the growth factor $D(z)$ and $\sigma(M)$, the dispersion of the smoothed matter field with a top-hat sphere of radius $R$ such that $M=\bar{\rho}_0(4\pi/3)R^3$. 

Several mass function models have been proposed \citep{Press_Schechter_1974, Sheth_Tormen_1999, Sheth_Tormen_2002, Jenkins_etal_2001, Warren_etal_2006, Tinker_etal_2008a, Bhattacharya_etal_2011}. Throughout this paper, we assume the universality of the mass function and adopt the model from \citet{Jenkins_etal_2001}, which gives
\begin{align}\label{for:massFct_Jenkins}
	f(\sigma) = 0.315 \exp\left( -\left|\ln\sigma\inv + 0.61\right|^{3.8} \right).
\end{align}

\subsection{Halo density profiles}
%\label{sect:basics_profiles}

We assume in this work that halos follow Navarro-Frenk-White (NFW) density profiles \citep{Navarro_etal_1996, Navarro_etal_1997}. The truncated version of these profiles is defined as
\begin{align}\label{for:NFW_profiles}
	\rho(r) = \frac{\rho_\rms}{(r/r_\rms) (1+r/r_\rms)^2}\Theta(r_\vir-r),
\end{align}
where $\Theta$ is the Heaviside step function. The NFW profiles are parameterized by two numbers: the central mass density $\rho_\rms$ and the scalar radius $r_\rms$. Depending on the convention, these two quantities can have different definitions. A universal way to express them is as follows:
\begin{align}
	r_\rms \equiv \frac{r_\Delta}{c}\ \ \text{and}\ \ \rho_\rms \equiv \rho_\mathrm{ref}\Delta\cdot\frac{1}{3}fc^3,
\end{align}
where $c$ is the concentration parameter and 
\begin{align}
	f \equiv \frac{1}{\ln(1+c)-c/(1+c)}.
\end{align}
Here, $\rho_\mathrm{ref}$ is the reference density, which may be the current critical density $\rho_{\crit,0}$, the critical density at $z$: $\rho_\crit(z)$, the current background density $\bar{\rho}_0$, or the background density at $z$: $\bar{\rho}(z)$. The factor $\Delta$ is the virial threshold above which halos are considered bound, which means that $M=\rho_\mathrm{ref}\Delta\cdot 4\pi r_\Delta^3/3$. This may be a redshift-dependent formula $\Delta_\vir(z)$, or a constant such as 200 or 500. In this paper, we adopt the definitions below:
\begin{align}
	r_\rms &\equiv \frac{r_\vir}{c} \label{for:r_s}\\
	\rho_\rms &\equiv \bar{\rho}(z)\Delta_\vir(z)\cdot\frac{1}{3}fc^3, \label{for:rho_s}
\end{align}
where $r_\vir$ is the physical virial radius and $\Delta_\vir(z)$ is a fitting function for a $w$CDM model, taken from Eqs. (16) and (17) from \citet{Weinberg_Kamionkowski_2003}.

The concentration parameter $c$ is redshift- and mass-dependent \citep{Bullock_etal_2001, Bartelmann_etal_2002, Dolag_etal_2004}. We use the expression proposed by \citet{Takada_Jain_2002}, which leads to
\begin{align}\label{for:M_C_relation}
	c(z,M) = \frac{c_0}{1+z}\left(\frac{M}{M_\star}\right)^{-\beta},
\end{align}
where the pivot mass $M_\star$ satisfies the condition $\delta_\rmc(z=0) = \sigma(M_\star)$, where $\delta_\rmc$ is the critical threshold for the spherical collapse model, given by Eq. (18) from \citet{Weinberg_Kamionkowski_2003}.

\subsection{Weak gravitational lensing}
%\label{sect:basics_lensing}

Consider a source to which the comoving radial distance from the observer is $w$. From the Newtonian potential $\phi$, one can derive the lensing potential $\psi$, following \citep[see, e.g.,][]{Schneider_etal_1998}
\begin{align}
	\psi(\btheta, w) \equiv \frac{2}{\rmc^2}\int_0^w \rmd w'\ \frac{f_K(w-w')}{f_K(w)f_K(w')}\ \phi\left( f_K(w')\btheta, w' \right),
\end{align}
where $\btheta$ is the coordinates of the line of sight, $f_K$ the comoving transverse distance, and $\rmc$ light speed. At the linear order, the lensing distortion is characterized by two quantities, the convergence $\kappa$ and the shear $\gamma_1+\rmi\gamma_2$, given by the second derivatives of $\psi$:
\begin{align}
	\kappa &\equiv \frac{1}{2}\left(\partial_1^2\psi + \partial_2^2\psi\right),\\
	\gamma_1 &\equiv \frac{1}{2}\left(\partial_1^2\psi - \partial_2^2\psi\right),\\
	\gamma_2 &\equiv \partial_1\partial_2\psi.
\end{align}
In other words, the linear distortion matrix $\mathcal{A}(\btheta)$, defined as $\mathcal{A}_{ij}(\btheta) = \delta_{ij} - \partial_i\partial_j\psi(\btheta)$ where $\delta_{ij}$ is the Kronecker delta, can be parametrized as
\begin{align}
	\mathcal{A}(\btheta) =
	\begin{pmatrix}
		1-\kappa - \gamma_1 &          - \gamma_2\\
		         - \gamma_2 & 1-\kappa + \gamma_1
	\end{pmatrix}.
\end{align}

Furthermore, the Newtonian potential is related to the matter density contrast $\delta$ via Poisson's equation in comoving coordinates:
\begin{align}
	\nabla^2\phi = \frac{3H^2_0 \OmegaM}{2a}\delta.
\end{align}
This provides an explicit expression of $\kappa$ as
\begin{align}\label{for:kappa_as_delta}
	\kappa(\btheta, w) = \frac{3H^2_0 \OmegaM}{2\rmc^2} \int_0^w \rmd w'\ \frac{f_K(w-w')f_K(w')}{f_K(w)} \frac{\delta\big( f_K(w')\btheta, w' \big)}{a(w')},
\end{align}
where $H_0$ is the Hubble parameter, $\OmegaM$ the matter density, and $a(w')$ the scale factor at the epoch to which the comoving distance from now is $w'$.

\newcommand{\cSqXSq}{\sqrt{c^2-x^2}}
\newcommand{\cSqOne}{\sqrt{c^2-1}}

The lensing signal contribution from halos with truncated NFW profiles is known. Defining $\theta_\rms=r_\rms/\DL$ as the ratio of the scalar radius to the angular diameter distance of the lens, if the density of the region not occupied by halos is assumed to be identical to the background, the convergence and the shear are given by computing the projected mass, which leads to
\begin{align}
	\kappa_\proj(\theta) = \sum_{\halo\rms} \kappa_\halo(\theta)\ \ \ \text{and}\ \ \ \gamma_\proj(\theta) = \sum_{\halo\rms} \gamma_\halo(\theta) \label{for:kappa_proj}
\end{align}
with
\begin{align}
	\kappa_\halo(\theta) = \frac{2\rho_\rms r_\rms}{\Sigma_\crit} G_\kappa\left(\frac{\theta}{\theta_\rms}\right)\ \ \ \text{and}\ \ \ \gamma_\halo(\theta) = \frac{2\rho_\rms r_\rms}{\Sigma_\crit} G_\gamma\left(\frac{\theta}{\theta_\rms}\right), \label{for:kappa_halo}
\end{align}
where $r_\rms$ and $\rho_\rms$ are respectively given by Eqs. \eqref{for:r_s} and \eqref{for:rho_s}, $\theta$ is the angular separation between the source and the center of the halo, and $\Sigma_\crit \equiv (\rmc^2/4\pi\rmG)(\DS/\DL\DLs)$ with $\rmG$ the gravitational constant, $\DS$ the angular diameter distance of the source, and $\DLs$ the angular diameter distance between the lens and the source. The dimensionless functions $G_\kappa$ and $G_\gamma$ are provided by \citet{Takada_Jain_2003a, Takada_Jain_2003b}. For computational reasons, it is useful to write $2\rho_\rms r_\rms=Mfc^2 / 2\pi r_\vir^2$, which can be obtained from Eqs. \eqref{for:r_s} and \eqref{for:rho_s}.

\subsection{Local noise level}
%\label{sect:methodology_noise}

In principle, the intrinsic ellipticity of galaxies can not be measured directly. We assume that both components of the ellipticity $\epsilon=\epsilon_1+\rmi\epsilon_2$ follow the same Gaussian distribution, such that its norm precisely follows a Rayleigh distribution. Since the ellipticity is bound by $\pm 1$, both distributions are truncated. We note $\sigma_\epsilon^2=\sigma_{\epsilon_1}^2+\sigma_{\epsilon_2}^2$ as the sum of the variances of both components. In this case, the noise for the smoothed convergence is also Gaussian, and its variance is given by \citep[see e.g.][]{VanWaerbeke_2000}
\begin{align}\label{for:global_noise}
	\sigma_\noise^2 = \frac{\sigma_\epsilon^2}{2n_\gala}\cdot\frac{\parallel W\parallel_2^2}{\parallel W\parallel_1^2},
\end{align}
where $n_\gala$ is the galaxy number density and $\parallel W\parallel_p$ stands for the $p$-norm of $W$ which is the smoothing kernel. The kernel does not need to be normalized because of the denominator in \for{for:global_noise}. For example, if $W$ is Gaussian with width $\theta_{\ker}$, $\parallel W\parallel_2^2/\parallel W\parallel_1^2 = 1/2\pi\theta_{\ker}^2$. For the starlet (see \sect{sect:filtering}), $\parallel W\parallel_2^2 = 5(2/5+5/63)^2 - 2(1/3+1/5+1/21+1/48)^2\approx 0.652^2$ can be solved analytically.

However, \for{for:global_noise} is the global noise level, which implies that sources are distributed regularly. In realistic conditions, random fluctuations, mask effects, and clustering of source galaxies can all lead to irregular distributions, which results in a non-constant noise level. To properly take this into account, we define the variance of the local noise as
\begin{align}\label{for:local_noise}
	\sigma_\noise^2(\btheta) = \frac{\sigma_\epsilon^2}{2}\cdot\frac{\sum_i W^2(\btheta_i-\btheta)}{\left(\sum_i |W(\btheta_i-\btheta)|\right)^2},
\end{align}
where $\btheta_i$ is the position of the $i$-th galaxy, and $i$ runs over all (non-masked) galaxies under the kernel $W$. Equation \eqref{for:local_noise} is also valid for the aperture mass (see next section), by replacing $W$ with $Q$ \citep{Schneider_1996}.

\section{Filtering}
\label{sect:filtering}

\subsection{Linear filters}
%\label{sect:filtering_linear}

In this work, we vary the filtering technique and study its impact on peak counts. Here, we present the linear filters $W$ used in this study. The description of the nonlinear technique can be found in \sect{sect:filtering_nonlinear_MRLens}. Let $\theta_{\ker}$ the size of the kernel and $x=\theta/\theta_{\ker}$. Then, the Gaussian smoothing kernel can be simply written as 
\begin{align}\label{for:Gaussian}
	W(x) \propto \exp\left(-x^2\right).
\end{align}

The second kernel that we study is the 2D \emph{starlet} function \citep{Starck_etal_2002}. It is defined as 
\begin{align}\label{for:starlet}
	W(x, y) = 4\phi(2x)\phi(2y) - \phi(x)\phi(y),
\end{align}
where $\phi$ is the B-spline of order 3, given by
\begin{align}
	\phi(x) = \frac{1}{12}\left(|x-2|^3 - 4|x-1|^3 + 6|x|^3 - 4|x+1|^3 + |x+2|^3\right).
\end{align}
Because of the property the B-spline, the starlet is a compensated function with compact support in $[-2,2]\times[-2,2]$. It does not conserve circular symmetry, but its isolines tend to be round. Since the starlet is compensated, it is similar to the $U$ function of the aperture mass, which is the last linear case that we consider.

The aperture mass $M_\ap$ can be obtained from all pairs of filters $(U, Q)$ such that (1) $U$ is circularly symmetric, (2) $U$ is a compensated function, and (3) filtering the convergence field with $U$ is equivalent to applying $Q$ to the tangential shear $\gamma_\rmt(\btheta=\theta\rme^{\rmi\varphi})\equiv -\gamma_1\cos(2\varphi)-\gamma_2\sin(2\varphi)$, where $\varphi$ is the complex angle of the source position with regard to the kernel center. With these conditions, convolving $\gamma_\rmt$ with $Q$ results in a filtered convergence map that is not affected by the mass-sheet degeneracy and the inversion problem. 

To satisfy the third condition, $Q$ has to be related to $U$ by
\begin{align}
	Q(\theta) \equiv \frac{2}{\theta^2}\int_0^\theta\rmd\theta'\ \theta'U(\theta')-U(\theta).
\end{align}
In this case, $M_\ap$ is given by
\begin{align}\label{for:M_ap}
	M_\ap(\btheta) \equiv \int\rmd^2\btheta'\ U(\btheta)\kappa(\btheta-\btheta') = \int\rmd^2\btheta'\ Q(\btheta)\gamma_\rmt(\btheta-\btheta').
\end{align}
Here, we are particularly interested in the $Q$ function proposed by \citet{Schirmer_etal_2004} and \citet{Hetterscheidt_etal_2005}, given by
\begin{align}\label{for:M_ap_Q}
	Q(x) \propto \frac{\tanh(x/x_\rmc)}{(x/x_\rmc)\left(1+\exp(a-bx)+\exp(-c+dx)\right)},
\end{align}
with $a=6$, $b=150$, $c=47$, $d=50$ to have a cutoff around $x=1$. This filter shape has been motivated by the tangential shear pattern generated by NFW halo profiles. Also, we set $x_\rmc=0.1$ as suggested by \citet{Hetterscheidt_etal_2005}. Note that $x=\theta/\theta_{\ker}$ is the distance to the center of the filter normalized by the kernel's size.

\subsection{A sparsity-based nonlinear filter}
%\label{sect:filtering_nonlinear}

In this section, we introduce a nonlinear filtering technique using the sparsity of signals.

\subsubsection{Sparse representation}
\label{sect:filtering_nonlinear_sparse}

In signal processing, a signal is sparse in a specific representation if most of the information is contained in only a few coefficients. This means that either only a finite number of coefficients is non zero, or the coefficients decrease fast when rank-ordered.

A straightforward example is the family of sine functions. In the real space, sine functions are not sparse. However, they are sparse in the Fourier space since they become the Dirac delta functions. More generally, periodic signals are sparse in the Fourier space.

Why is this interesting? Because white noise is not sparse in any representation. Therefore, if the information of the signal can be compressed into a few strong coefficients, it can easily be separated from the noise. This concept of sparsity has been widely used in the signal processing domain for applications such as denoising, inpainting, deconvolution, inverse problem, or other optimization problems \citep{Daubechies_etal_2004, Candes_Tao_2006, Elad_Aharon_2006, Candes_etal_2008, Fadili_etal_2009}. Examples can also be found for studying astophysical signals \citep{Lambert_etal_2006, Pires_etal_2009, Bourguignon_etal_2011, Carrillo_etal_2012, Bobin_etal_2014, NgoleMboula_etal_2015, Lanusse_etal_2016}.

\subsubsection{Wavelet transform}
%\label{sect:filtering_nonlinear_wavelet}

From the previous section, one can see that the sparsity of a signal depends on its representation basis. In which basis is the weak lensing signal sparse? A promising candidate is the wavelet transform which decomposes the signal into a family of scaled and translated functions. Wavelet functions are all functions $\psi$ which satisfy the \emph{admissibility condition}:
\begin{align}
	\int_0^{+\infty} |\hat{\psi}(k)|^2 \frac{\rmd k}{k} < +\infty.
\end{align}
One of the properties implied by this condition is $\int\psi(x)\rmd x=0$, which restricts $\psi$ to a compensated function. In other words, one can consider wavelet functions as highly localized functions with a zero mean. Such a function $\psi$ is called the mother wavelet, which can generate a family of daughter wavelets such as
\begin{align}
	\psi_{a,b}(x) = \frac{1}{\sqrt{a}}\psi\left( \frac{x-b}{a} \right),
\end{align}
which are scaled and translated versions of the mother $\psi$.

\begin{figure}[tb]
	\centering
	\includegraphics[width=\columnwidth]{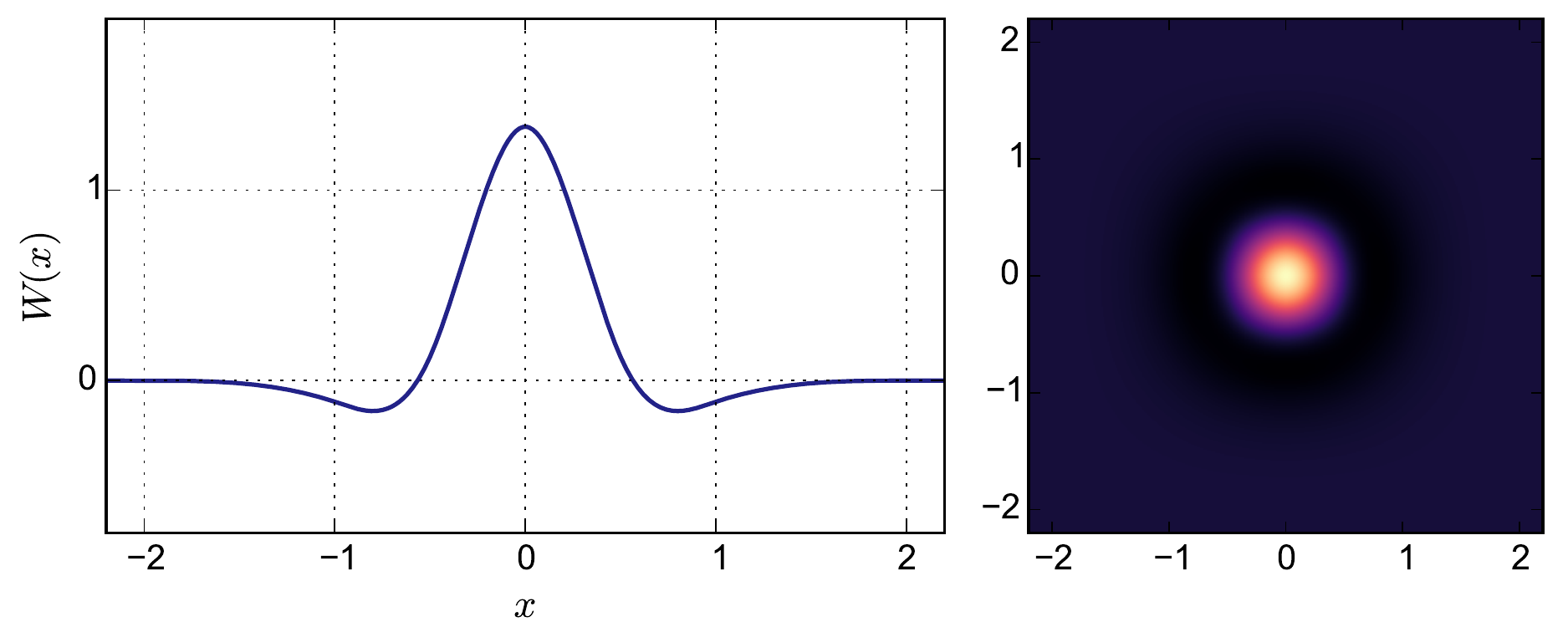}
	\caption{\emph{Left panel}: the profile of the 2D starlet. It has a finite support [-2, 2]. \emph{Right panel}: the bird-eye view of the 2D starlet.}
	\label{fig:Starlet}
\end{figure}

The wavelet transform \citep[see e.g. Chaps. 2 and 3 of][]{Starck_etal_2002} refers to the decomposition of an input image into several ones of the same size each associated to a specific scale. Due to the property of wavelet functions, each resulting image gives the details of the original one at different scales. If we stack all the images, we recover the original signal.

In the peak-count scenario, peaks which are generated by massive clusters are considered as signals. Like clusters, these signals are local point-like features, and therefore have a sparse representation in the wavelet domain. As described in \sect{sect:filtering_nonlinear_sparse}, white noise is not sparse. So one simple way to reduce the noise is to transform the input image into the wavelet domain, set a relatively high threshold $\lambda$, cut out weak coefficients smaller than $\lambda$, and reconstruct the clean image by stacking the thresholded images. In this paper, we use the 2D starlet function as the mother wavelet, given by \for{for:starlet}, which satisfies the admissibility condition. As shown by \fig{fig:Starlet}, it highlights round features as we assume for dark matter halos.

\subsubsection{The MRLens filter}
\label{sect:filtering_nonlinear_MRLens}

In this study, we apply the nonlinear filtering technique \emph{MultiResolution tools for gravitational Lensing} \citep[\MRLens,][]{Starck_etal_2006} to lensing maps. \MRLens\ is an iterative filtering based on Bayesian framework that uses a multiscale entropy prior and the false discovery rate (FDR, \citealt{Benjamini_Hochberg_1995}) which allows to derive robust detection levels in wavelet space.

More precisely, \MRLens\ first applies a wavelet transform to a noisy map. Then, in the wavelet domain, it determines the threshold by FDR. The denoising problem is regularized using a multiscale entropy prior only on the non-significant wavelet coefficients. Readers are welcome to read \citet{Starck_etal_2006} for a detailed description of the method.

Note that, whereas \citet{Pires_etal_2009a} selected peaks from different scales separately before final reconstruction, in this paper, we count peaks on the final reconstructed map. In fact, the methodology of \citet{Pires_etal_2009a} is close to filtering with a lower cutoff in the histogram defined by FDR, thus similar to starlet filtering. With the vocabulary defined in \sect{sect:intro}, \citet{Pires_etal_2009a} followed the separated strategy and here we attempt the combined strategy. This choice provides a comparison between cosmological information extracted with two strategies, by comparing starlet filtering to the \MRLens\ case.

\section{Methodology}
\label{sect:methodology}

\defcitealias{Lin_Kilbinger_2015a}{II}
In this section, we review our peak count model, and detail the improvements \PaperI\ and \PaperII\ that we introduce here.

\subsection{General concept of our model}
%\label{sect:methodology_concept}

\begin{figure}[tb]
	\centering
	\includegraphics[width=\columnwidth]{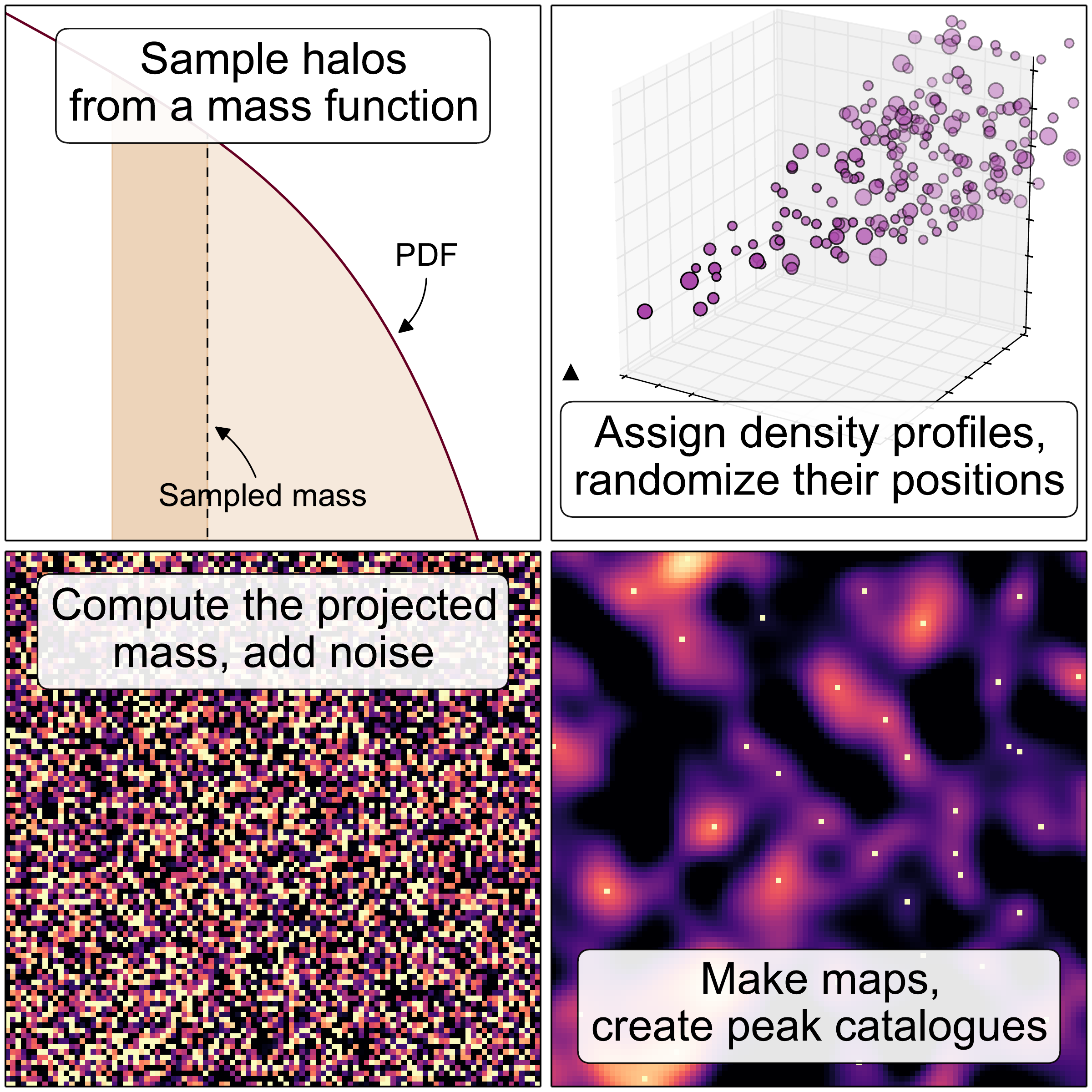}
	\caption{Illustration of our model in four panels.}
	\label{fig:LK_model_illustration_1}
\end{figure}

In \PaperI\ and \PaperII, we proposed a fast stochastic model for predicting weak lensing peak counts. The general concept is to bypass the complex and time-consuming $N$-body process. Our model generates ``fast simulations'' based on halo sampling, and counts peaks from lensing maps obtained from these simulation boxes, as illustrated in \fig{fig:LK_model_illustration_1}.
\defcitealias{Lin_Kilbinger_2015a}{Paper II}

To achieve this, we made two major assumptions. First, diffuse matter was considered to contribute little to peak counts. Second, we supposed that halo correlation had a minor impact. In \PaperI, we found that neither of them can be neglected alone. However, combining these two assumptions yielded a good approximation for the peak count prediction. 

The advantages provided by our model can be characterized by three properties: they are fast, flexible, and they provide the full PDF information. First, sampling from the mass function is very efficient. It requires about ten seconds for creating a 36-deg$^2$ field on a single-CPU computer. Second, our model is flexible because survey-related properties, such as masking and realistic photo-$z$ errors, can be included in a straightforward way thanks to its forward nature. Third, because of the stochasticity, the PDF of the observables is available. As we showed in \PaperII, this PDF information allows us not only to estimate the covariance matrix, but also to use other more sophisticated inference methods, such as approximate Bayesian computation.

Our model has been implemented in the language C as the software \textsc{Camelus}, which is available on GitHub \footnote{\url{http:github.com/Linc-tw/camelus}}.

\subsection{Settings for the pipeline: from the mass function to peak catalogs}
%\label{sect:methodology_pipeline}

In this section, we explain in detail how peak counts are generated from an initial cosmological model. We first sampled halos from \for{for:massFct_Jenkins}. The sampling range was set to $M=[5\dixx{12}, 10^{17}]\ \Msol/h$. This was done for 30 equal redshift bins from 0 to 3, on a field adequately larger than 36 deg$^2$ so that border effects were properly eliminated. For each bin, we estimated the volume of the slice, the mass contained in the volume and in the sampling range, such that the total mass of the samples corresponded to this value. Then, these halos were placed randomly and associated with truncated NFW profiles using \for{for:NFW_profiles}, where the mass-concentration relation was given by \for{for:M_C_relation}. We note that studying the impact of the mass function modeling or profile modeling with our model is possible. Nevertheless, this is not the aim of this paper.

\defcitealias{Lin_Kilbinger_2015a}{II}
We extended our model from \PaperI\ and \PaperII\ to include realistic observing properties as follows. First, we considered a realistic redshift distribution of sources for the analysis. We assumed a gamma distribution following \citet{Efstathiou_etal_1991}
\begin{align}
	p(z) = \frac{z^2}{2z^3_0}\exp\left(-\frac{z}{z_0}\right),
\end{align}
where $z_0=0.5$ is the pivot redshift value. The positions of sources were random. We set the source number density to $n_\gala=12\ \arcmin\invSq$, which corresponded to a CFHTLenS-like survey \citep{Heymans_etal_2012}. The intrinsic ellipticity dispersion was $\sigma_\epsilon=0.4$, which is also close to the CFHTLenS survey \citep{Kilbinger_etal_2013}.
\defcitealias{Lin_Kilbinger_2015a}{Paper II}

Second, we considered masks in our model. We applied the same characteristic mask to each of the realizations of our model. This mask was taken from the W1 field of CFHTLenS.

For each galaxy, we computed $\kappa_\proj$ and $\gamma_\proj$ using Eqs. \eqref{for:kappa_proj} and \eqref{for:kappa_halo}. However, as we already evoked in \PaperI, $\kappa_\proj$ can not be considered as the true convergence since it is always positive. In fact, \for{for:kappa_proj} can be derived from \for{for:kappa_as_delta} by replacing the density contrast $\delta$ with $\rho/\bar{\rho}$, thus it becomes positive. To handle this difference, we subtracted the mean of the field $\overline{\kappa}_\proj$ from $\kappa_\proj$. This subtraction is supported by $N$-body simulations. For example, for simulations used in \PaperI, we found $\bar{\kappa}\sim 8\dixd{-4}$, that implied that the mean almost vanished. Finally, we computed the observed ellipticity as $\epsilon\upp{\rmo}=g_\proj+\epsilon\src$, where $g_\proj\equiv\gamma_\proj/(1-(\kappa_\proj-\overline{\kappa}_\proj))$ is the reduced shear and $\epsilon\src$ is the intrinsic ellipticity. 

Comparing different mass mapping techniques is the subject of this study. We tested the Gaussian kernel, the starlet function, the aperture mass with the hyperbolic tangent function, and the nonlinear filtering technique \MRLens\ in our model. Except for the aperture mass, we first binned galaxies into map pixels and took the mean of $\epsilon\upp{\rmo}$ as the pixel's value for the reason of efficiency. The pixel size was 0.8 arcmin. This resulted in regularly spaced data so that the algorithm can be accelerated. Then, the Kaiser-Squires inversion \citep[KS inversion,][]{Kaiser_Squires_1993} was used before filtering. We did not correct for the reduced shear, for example by iteratively using the KS inversion, since the linear inversion conserves the original noise spectrum and produces less artefacts. By applying exactly the same processing to both observation and model prediction, we expect the systematics related to inversion (e.g., boundary effects, missing data, and negative mass density) to be similar so that the comparison is unbiased. For the aperture mass, the pixel's value was evaluated by convolving directly the lensing catalog with the $Q$ filter (Eq. \ref{for:M_ap}), successively placed at the center of each pixel \citep[see also][]{Marian_etal_2012, Martinet_etal_2015}. The choice of filter sizes is detailed in \sect{sect:methodology_filters}.

Because of the presence of masks, we selected peaks based on the concept of the filling factor $f(\btheta)$ (\citealt{VanWaerbeke_etal_2013}; \citetalias{Liu_etal_2015a} \citeyear{Liu_etal_2015a}). A local maximum was selected as a peak only if $f(\btheta)\geq \lambda \bar{f}$, where $\bar{f}$ is the mean of $f$ over the map. We set $\lambda=0.5$. For analyses using binning, the filling factor was simply defined as the number of galaxies $N(\btheta)$ inside the pixel at $\btheta$. For the aperture mass, it was the $Q$-weighted sum of the number counts. In other words,
\begin{align}
	f(\btheta) \equiv \left\{\begin{array}{ll}
		N(\btheta)          & \text{if galaxies are binned,}\\
		\sum_i Q(\btheta_i) & \text{for the aperture mass,} 
	\end{array}\right.
\end{align}
where $\btheta_i$ is the position of the $i$-th galaxy.
 
Furthermore, peaks were selected based on their local noise level. For linear filters (the Gaussian, the starlet, the aperture mass), the local noise level was determined by \for{for:local_noise}. The height of peaks $\nu$ was then defined as the S/N by
\begin{align}
	\nu(\btheta) \equiv \left\{\begin{array}{ll}
		(\kappa\ast W)(\btheta)/\sigma_\noise(\btheta) & \text{if Gaussian or starlet,}\\
		M_\ap(\btheta)/\sigma_\noise(\btheta) & \text{if aperture mass,} 
	\end{array}\right.
\end{align}
where $\ast$ is the convolution operator. However, for the nonlinear technique, the noise after filtering is not Gaussian anymore. The so-called noise level cannot be properly defined. In this case, we simply selected peaks on $\kappa$.

\subsection{Settings for filters and data vectors}
\label{sect:methodology_filters}

\begin{figure}[tb]
	\centering
	\includegraphics[width=\columnwidth]{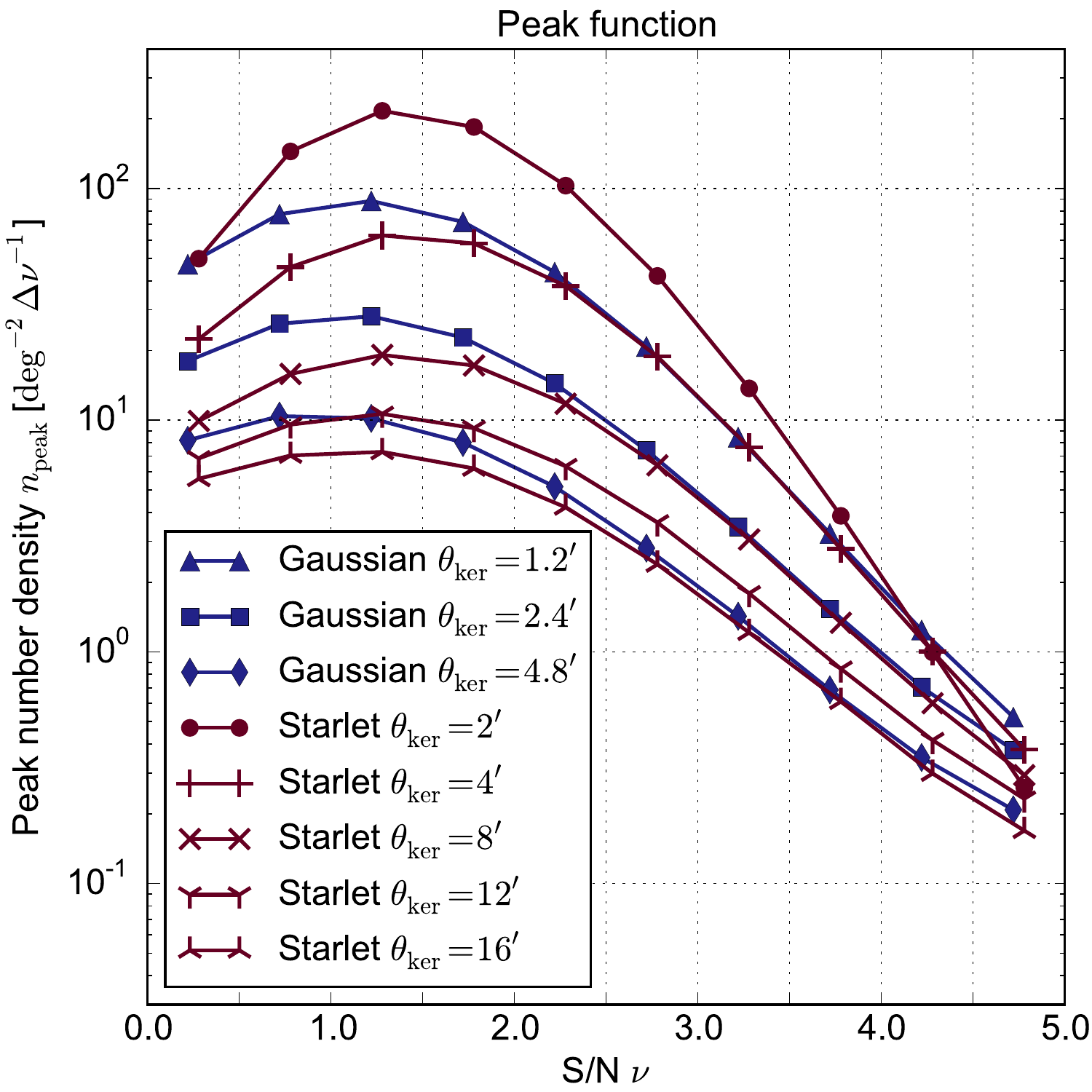}
	\caption{Peak function for different kernel sizes for an input cosmology $(\OmegaM, \sigEig, \wZero)=(0.28, 0.82, -0.96)$. The number counts are the mean over 400 realizations of 36 deg$^2$.}
	\label{fig:peakHist_gauss_vs_star}
\end{figure}

\begin{table}
	\centering
	\caption{List of kernel sizes $\theta_{\ker}$. We choose $\theta_{\ker}$ based on $\sigma_\noise$ such that the corresponding values are similar. The quantity $\sigma_\noise$ is computed using \for{for:global_noise} with $n_\gala=12~\arcmin\invSq$ and $\sigma_\epsilon=0.4$.}
	\begin{tabular}{rccccc}
		\hline\hline\\[-1.8ex]
		Kernel                   & \multicolumn{5}{l}{Gaussian}\\
		$\theta_{\ker}$ [arcmin] & 1.2   & 2.4   & 4.8    & & \\
		$\sigma_\noise$          & 0.027 & 0.014 & 0.0068 & & \\
		\hline\\[-1.8ex]
		Kernel                   & \multicolumn{5}{l}{Starlet}\\
		$\theta_{\ker}$ [arcmin] & 2     & 4     & 8      & 12     & 16\\
		$\sigma_\noise$          & 0.027 & 0.014 & 0.0068 & 0.0045 & 0.0034\\
		\hline\\[-1.8ex]
		Kernel                   & \multicolumn{5}{l}{$M_\ap$ $\tanh$}\\
		$\theta_{\ker}$ [arcmin] & 2.125 & 4.25  & 8.5    & & \\
		$\sigma_\noise$          & 0.027 & 0.014 & 0.0068 & & \\
		\hline
	\end{tabular}
	\label{tab:scales}
\end{table}

The aim of this paper is to compare the performance of linear and nonlinear filters for peak counts. The linear filters were parametrized with a single parameter, which is the size of the kernel $\theta_{\ker}$. We proposed two possible solutions for comparing between kernels of different shape. The first was to choose $\theta_{\ker}$ such that the 2-norms have the same value if kernels are normalized (by their respective 1-norms). The reason for this is that if the ratio of the 2-norm to the 1-norm is identical, then the comparison is based on the same global noise level (Eq. \ref{for:global_noise}). \tab{tab:scales} shows various values of $\theta_{\ker}$ that we used in this studies and the corresponding $\sigma_\noise$ for different linear filters. The second way was to calculate peak-count histograms, and set $\theta_{\ker}$ such that peak abundance was similar. \figFull{fig:peakHist_gauss_vs_star} shows an example for the Gaussian and starlet kernels with $\theta_{\ker}$ taken from \tab{tab:scales}. We observe that, for Gaussian filtering with $\theta_{\ker}=$ 1.2, 2.4, and 4.8 arcmin, the correspondence for starlet filtering based on peak abundance is $\theta_{\ker}=$ 4, 8, and 16 arcmin if we focus on peaks with $2.5\leq\nu\leq4.5$, while the correspondence based on the noise level is $\theta_{\ker}=$ 2, 4, and 8 arcmin. In \sect{sect:results:filtering}, we will examine both comparison methods.

The data vector $\bx$, for linear filters, was defined as the concatenation of several S/N histograms. In \PaperII, we have found that the number counts from histograms are the most appropriate form to derive cosmological information from peak counts. After testing several values of $\nu_\minn$, we only kept peaks above $\nu_\minn=1$ for each kernel size. This choice maximized the figure of merit of parameter constraints. Thus, we reconfirm that ignoring peaks with $\nu\leq 3$ corresponds to a loss of cosmological information \citep{Yang_etal_2013}. Peaks were then binned with width of $\Delta\nu=0.5$ up to $\nu=5$, and the last bin was $[5,+\infty[$ for each scale. For each $\bx$, the effective field size from which peaks were selected was 6 $\times$ 6 deg$^2$. Border effects were mitigated by taking adequately larger fields for halos and galaxies. The pixel size was 0.8 arcmin, so a map contained 450 $\times$ 450 pixels.

For the nonlinear filter, the notion of noise level does not easily apply. In fact, to determine the significance of a rare event from any distribution, instead of using the emperical standard deviation, it is more rigourous to obtain first the $p$-value and find how much $\sigma$ this value is associated with if the distribution was Gaussian. However, even if we compute the standard deviation instead, this process is still too expensive computationally for our purpose. Therefore, we bin peaks directly by their $\kappa$ values into [0.02, 0.03, 0.04, 0.06, 0.10, 0.16, $+\infty$[. This configuration was chosen such that the average count per bin is large enough to assume a Gaussian fluctuation.

\subsection{Sampling in the parameter space}
%\label{sect:methodology_sampling}

In this paper, we considered a three-dimensional parameter space, constructed with $(\OmegaM, \sigEig, \wZero)$, where $\wZero$ is the constant term of the equation of state of the dark energy. The values of other cosmological parameters were $h=0.78$, $\OmegaB=0.047$, and $n_\rms=0.95$. We assumed a flat Universe. The mock observation was generated by a realization of our model, using a particular set $(\OmegaM, \sigEig, \wZero)=(0.28, 0.82, -0.96)$ as input parameters. In this way, we only focus on the precision of our model.

\begin{figure}[tb]
	\centering
	\includegraphics[width=0.75\columnwidth]{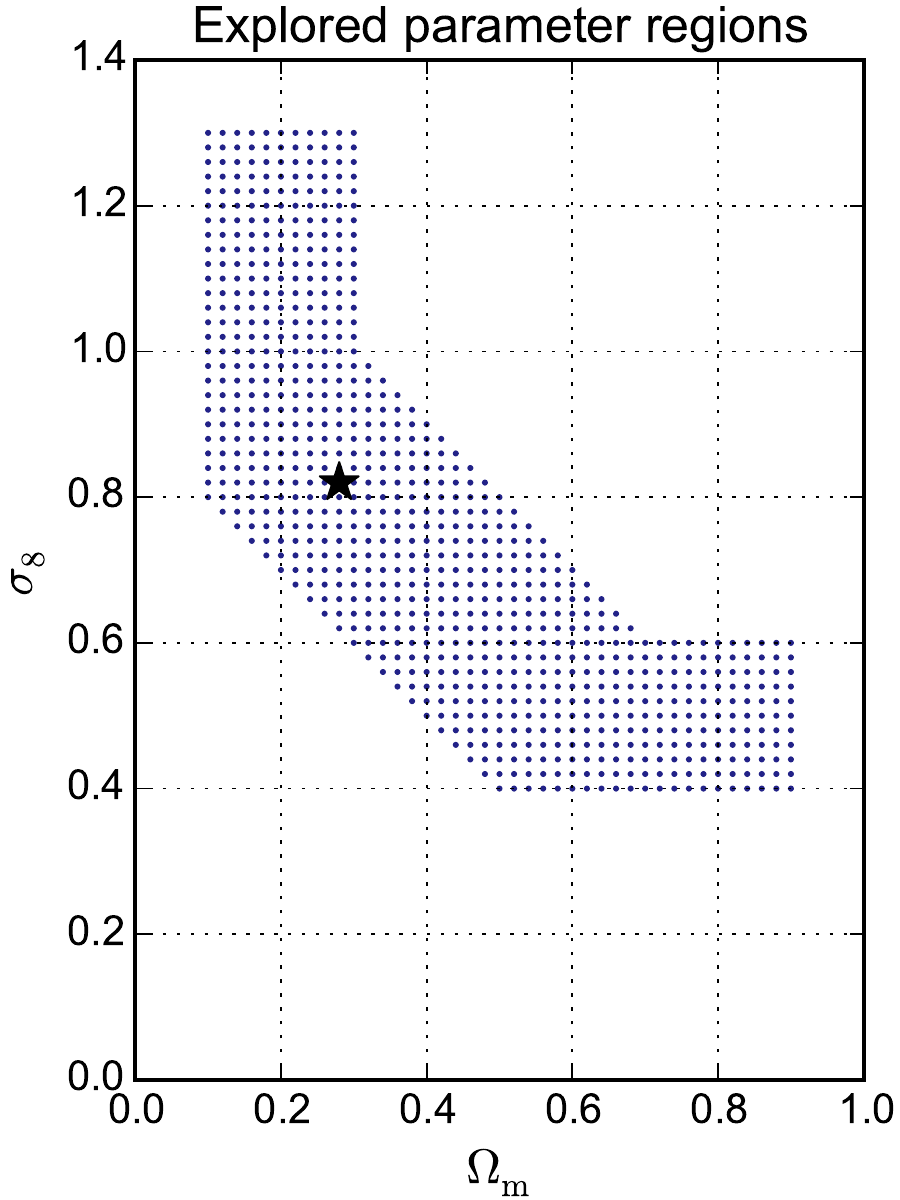}
	\caption{Distribution of evaluated parameter points on the $\OmegaM$-$\sigEig$ plane. This figure can be considered as a slice of points with the same $\wZero$. There are in total 46 slices of 816 points.}
	\label{fig:Explored_regions}
\end{figure}

\begin{figure*}[tb]
	\centering
	\includegraphics[width=17cm]{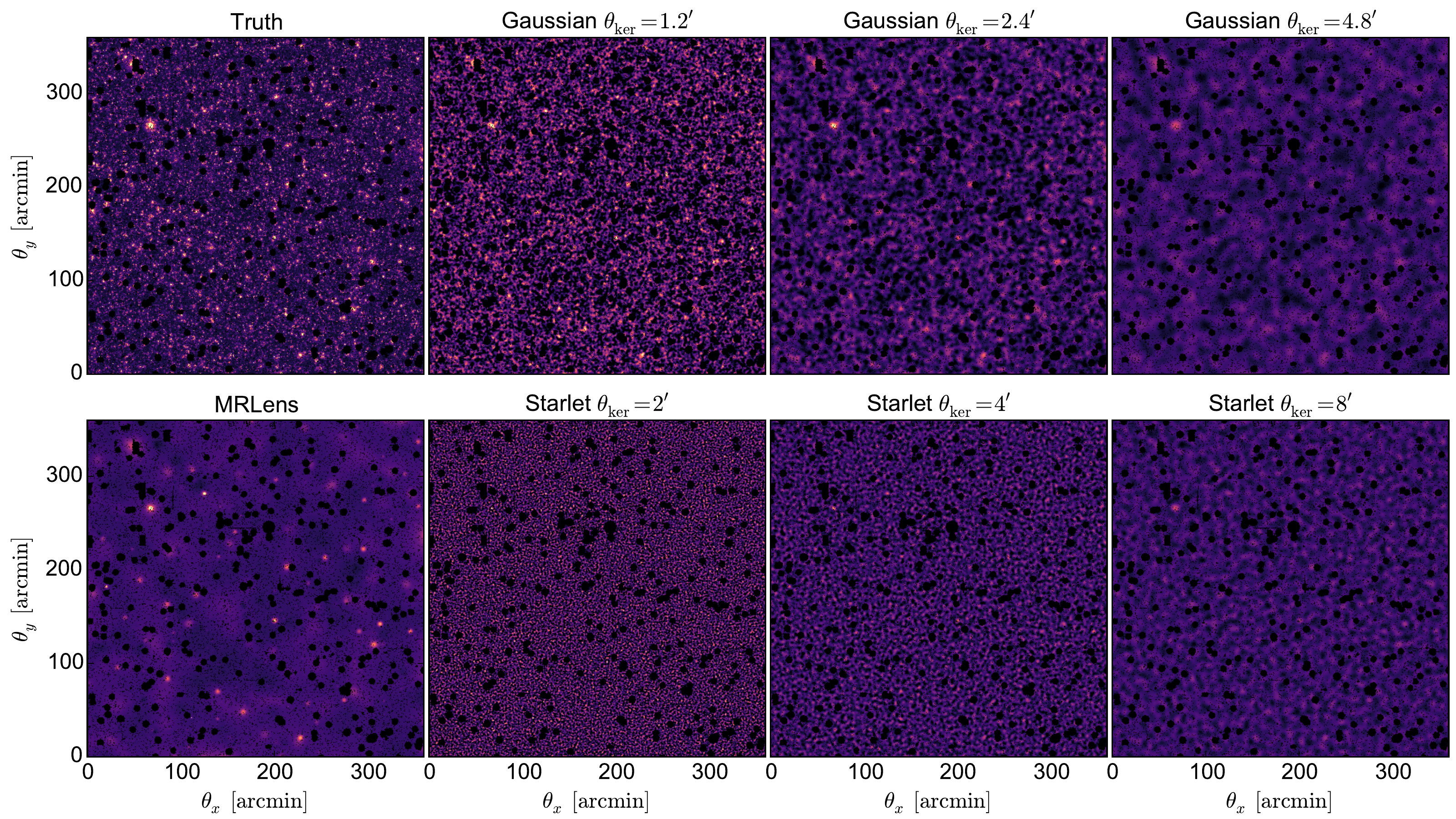}
	\caption{Maps taken from one of the simulations. The true map is made by calculating $\kappa_\proj$ without noise. The panels of the rest are different filtering techniques applied on the map obtained from a KS inversion after calculating $\epsilon\upp{\rmo}=g_\proj+\epsilon\src$. The black areas are masks. The unit of kernel sizes is arcmin.}
	\label{fig:allFilters}
\end{figure*}

We processed simulation runs in two different ways. The first one consists of interpolating the likelihood, from which we draw credible regions from Bayesian inference, and the second is approximate Bayesian computation. Both approaches are explained in the following sections.

\subsubsection{Copula likelihood}
%\label{sect:methodology_sampling_copula}

The copula likelihood comes from the copula transform which is a series of 1D transformations, which turn the marginals of a multivariate distribution into the desired target functions. In other words, it corresponds to applying successive changes of variables to a multivariate distribution. According to Sklar’s theorem \citep{Sklar_1959}, these transformations always exist. One may be interested in specific transformations such that in the new space, all marginals of the studied distribution are Gaussian. Then, the joint distribution in the new variables is closer to Gaussian in most cases. By combining the transformations mentioned above with the Gaussian likelihood, one obtains the copula likelihood.

We use the copula likelihood with covariances varying with cosmology. Let $d$ be the dimension of the data vector. Given a parameter set $\bpi$, for all $i=1,\ldots,d$, we note $x^\model_i(\bpi)$ as the $i$-th component of the model prediction, $\hat{\sigma}_i(\bpi)$ as the corresponding dispersion, and $\hat{P}_i(\cdot|\bpi)$ the $i$-th initial marginal. We also note $\bx^\obs$, $\widehat{\bC}$ and $\widehat{\bC\inv}$ as the observed data vector, the estimated covariance, and its inverse, respectively. The copula log-likelihood is
\begin{align}
	L &\equiv \ln\left[\det \widehat{\bC}(\bpi)\right] \notag\\
	&+ \sum_{i=1}^d \sum_{j=1}^d \left(q_i^\obs(\bpi) - x^\model_i(\bpi)\right) \widehat{C\inv_{ij}}(\bpi) \left(q_j^\obs(\bpi) - x^\model_j(\bpi)\right) \notag\\
	&- 2 \sum_{i=1}^d \ln\hat{\sigma}_i(\bpi) - \sum_{i=1}^d \left(\frac{q_i^\obs(\bpi) - x^\model_i(\bpi)}{\hat{\sigma}_i(\bpi)}\right)^2 \notag\\
	&- 2 \sum_{i=1}^d \ln\hat{P}_i(x^\obs_i|\bpi),\label{for:likelihood}
\end{align}
where $q^\obs_i(\bpi)$ is such that $\Phi_i(q^\obs_i)=\hat{F}_i(x^\obs_i - x^\model_i)$, knowing that $x^\model_i$ is the model prediction, $\hat{F}_i$ is the cumulative distribution of $\hat{P}_i$, and $\Phi_i$ is the cumulative of the normal distribution with the same mean and variance as $\hat{P}_i$. A more detailed description and the derivation of the copula can be found in Sect. 4 of \PaperII.

All the quantities required by the copula likelihood are provided by our model. Consider a set of $N$ model realizations. Denoting $x_i\upp{k}$ as the $i$-th component of the $k$-th realization, we use
\begin{align}
	x_i^\model   &= \frac{1}{N}\sum_{k=1}^N x_i\upp{k}, \label{for:estimator_mean}\\
	\hat{C}_{ij} &= \frac{1}{N-1}\sum_{k=1}^N \left(x_i\upp{k}-x_i^\model\right) \left(x_j\upp{k}-x_j^\model\right), \label{for:estimator_cov}\\
	\widehat{\bC\inv} &= \frac{N-d-2}{N-1}\ \widehat{\bC}\inv, \ \ \text{and} \label{for:estimator_invCov}\\
	\hat{P}_i(x_i) &= \frac{1}{N}\sum_{k=1}^N \frac{1}{h_i}W\left(\frac{x_i-x_i\upp{k}}{h_i}\right) \label{for:estimator_marginal}
\end{align}
for the estimations, where $d$ is the dimension of $\bx$, $W$ is the Gaussian kernel, and $h_i=(4/3N)^{1/5}\hat{\sigma}_i$. Note that the model prediction $\bx^\model$ is nothing but the average over the realization set; the inverse covariance matrix is unbiased \citep{Hartlap_etal_2007} to good accuracy (see also \citealt{Sellentin_Heavens_2016}); and \for{for:estimator_marginal} is a kernel density estimation (KDE).

We evaluated the copula likelihoods, given by \for{for:likelihood}, on a grid. The range of $\wZero$ is [-1.8, 0], with $\Delta\wZero=0.04$. Concerning $\OmegaM$ and $\sigEig$, only some particular values were chosen for evaluation in order to reduce the computing cost. This resulted in 816 points in the $\OmegaM$-$\sigEig$ plane, as displayed in \fig{fig:Explored_regions}, and the total number of parameter sets was 37536. For each parameter set, we carried out $N=400$ realizations of our model, to estimate $L$ using Eqs. \eqref{for:likelihood}, \eqref{for:estimator_mean}, \eqref{for:estimator_cov}, \eqref{for:estimator_invCov}, and \eqref{for:estimator_marginal}. Each realization produced data vectors for three cases: (1) the Gaussian kernel, (2) the starlet kernel, (3) \MRLens, so that the comparisons between cases are based on the same stochasticity. The aperture mass was not included here because of the time consuming convolution of the unbinned shear catalog with the filter $Q$. The FDR $\alpha$ of \MRLens\ was set to 0.05. A map example is displayed in \fig{fig:allFilters} for the three cases and the input simulated $\kappa$ field.

\begin{table}
	\centering
	\caption{Definition of the data vector $\bx$ for PMC ABC runs. The 9 bins of $\nu$ are [1, 1.5, 2, $\ldots$, 4, 4.5, 5, $+\infty$[, and the 6 bins of $\kappa$ are [0.02, 0.03, 0.04, 0.06, 0.10, 0.16, $+\infty$[. The symbol $d$ is the total dimension of $\bx$, and $\alpha$ stands for the input value of FDR for \MRLens.}
	\begin{tabular}{clcc}
		\hline\hline\\[-2ex]
		Filter          & $\theta_{\ker}$ [arcmin] or $\alpha$ & Number of bins  & $d$\\
		\hline\\[-1.8ex]
		Gaussian        & $\theta_{\ker}=$ 1.2, 2.4, 4.8       & 9 $\nu$ bins    & 27\\
		Starlet         & $\theta_{\ker}=$ 2, 4, 8             & 9 $\nu$ bins    & 27\\
		$M_\ap$ $\tanh$ & $\theta_{\ker}=$ 2.125, 4.25, 8.5    & 9 $\nu$ bins    & 27\\
		\MRLens         & $\alpha=0.05$                        & 6 $\kappa$ bins & 6\\
		\hline
	\end{tabular}
	\label{tab:x_mod_ABC}
\end{table}

\subsubsection{Population Monte Carlo approximate Bayesian computation}
%\label{sect:methodology_sampling_ABC}

The second analysis adopts the approximate Bayesian computation (ABC) technique. ABC bypasses the likelihood evaluation to estimate directly the posterior by accept-reject sampling. It is fast and robust, and has already had several applications in astrophysics (\citealt{Cameron_Pettitt_2012}; \citealt{Weyant_etal_2013}; \citealt{Robin_etal_2014}; \PaperII; \citealt{Killedar_etal_2015}). Here, we use the Population Monte Carlo ABC (PMC ABC) algorithm to constrain parameters. This algorithm adjusts the tolerance level iteratively, such that ABC posterior converges. A detailed description of the PMC ABC algorithm can be found in Sect. 6 of \PaperII. 

We ran PMC ABC for four cases: the Gaussian kernel, the starlet kernel, the aperture mass with the hyperbolic tangent function, and \MRLens\ with $\alpha=0.05$. For the three first linear cases, the data vector $\bx$ was composed of three scales. The S/N bins of each scale were [1, 1.5, 2, $\ldots$, 4, 4.5, 5, $+\infty$[, which result in 27 bins in total (\tab{tab:x_mod_ABC}). For \MRLens, $\bx$ was a 6-bin $\kappa$ histogram, which is the same as for the analysis using the likelihood. 

Concerning the ABC parameters, we used 1500 particles in the PMC process. The iteration stoped when the success ratio of accept-reject processes fell below 1\%. Finally, we tested two distances. Between the sampled data vector $\bx$ and the observed one, $\bx^\obs$, we considered a simplified distance $D_1$ and a fully correlated one $D_2$, which are respectively defined as
\begin{align}
	D_1\left(\bx, \bx^\obs\right) &\equiv \sqrt{\sum_i\frac{\left(x_i-x_i^\obs\right)^2}{C_{ii}}}, \label{for:D_1}\\
	D_2\left(\bx, \bx^\obs\right) &\equiv \sqrt{\left(\bx-\bx^\obs\right)^T \bC\inv\left(\bx-\bx^\obs\right)}, \label{for:D_2}
\end{align}
where $C_{ii}$ and $\bC\inv$ are now independent from cosmology, estimated using Eqs. \eqref{for:estimator_cov} and \eqref{for:estimator_invCov} under $(\OmegaM, \sigEig, \wZero)=(0.28, 0.82, -0.96)$. Note that $D_1$ has been shown in \PaperII\ to be able to produce constraints which agree well with the likelihood. However, with multiscale data, bins could be highly correlated, and therefore we also ran ABC with $D_2$ in this paper.

\section{Results}
\label{sect:results}

\subsection{Comparing filtering techniques using the likelihood}
\label{sect:results:filtering}

We propose two methods to measure the quality of constraints. The first indicator is the uncertainty on the derived parameter $\Sigma_8$. Here, we define $\Sigma_8$ differently from the literature:
\begin{align}\label{for:DCSE}
	\Sigma_8 \equiv \left(\frac{\OmegaM+\beta}{1-\alpha}\right)^{1-\alpha} \left(\frac{\sigma_8}{\alpha}\right)^{\alpha}.
\end{align}
The motivation for this definition is to measure the contour width independently from $\alpha$. With the common definition $\Sigma_8 \equiv \sigma_8(\OmegaM/\text{pivot})^\alpha$, the variation on $\Sigma_8$ under different $\alpha$ does not correspond to the same width on the $\OmegaM$-$\sigEig$ plane. The 1-$\sigma$ error bar on $\Sigma_8$, $\DCSE$, is obtained using the same method as in \PaperII. The second indicator is the FoM defined as the inverse of the 2-$\sigma$ contour area for $\OmegaM$ and $\sigEig$. 

First, we test the maximum information that Gaussian kernels can extract. \tab{tab:FoM_likelihood} shows the FoM from the marginalized likelihood. We can see that adding $\theta_{\ker}=$ 2.4 and 4.8 arcmin to the filter with 1.2 arcmin has no siginificant effect on constraints. The constraints from the smallest filter are the most dominant ones among all.

\begin{figure*}[tb]
	\centering
	\includegraphics[width=8.5cm]{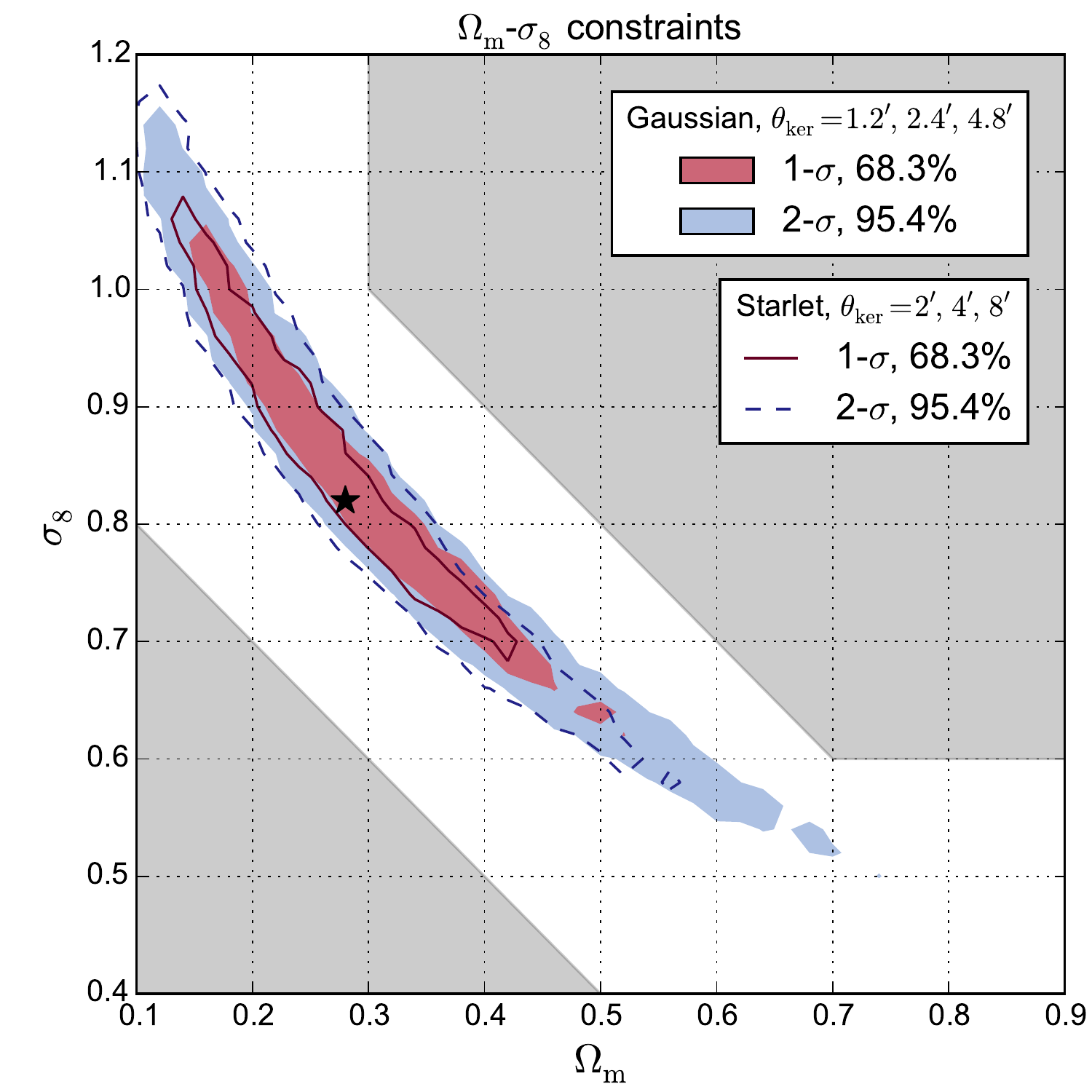}
	\includegraphics[width=8.5cm]{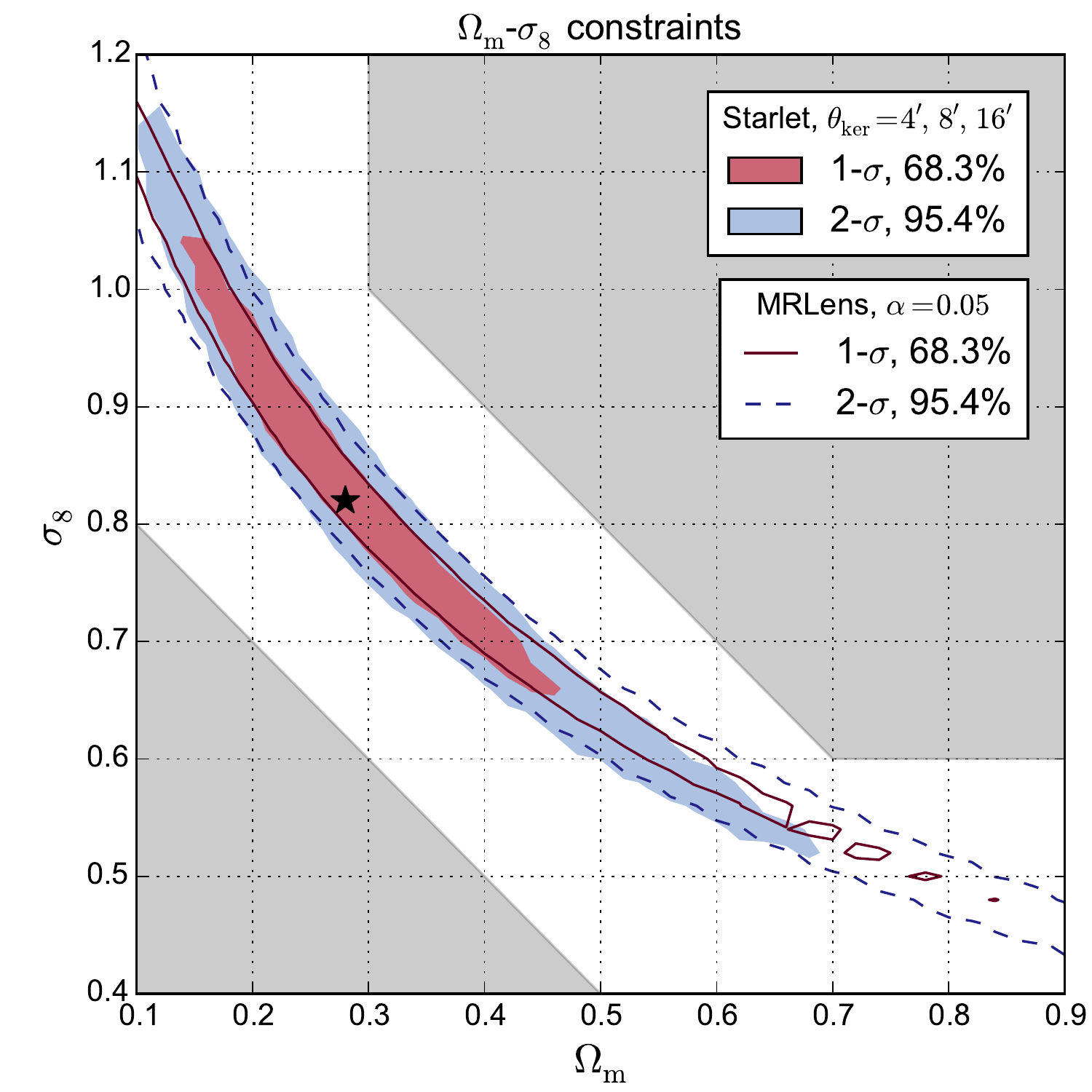}
	\caption{$\OmegaM$-$\sigEig$ constraints from four different cases. \textit{Left panel}: the Gaussian case (colored regions) and the starlet case with three corresponding scales based on the noise level (solid and dashed contours, $\theta_{\ker}=$ 2, 4, and 8 arcmin). \textit{Right panel}: the starlet case with three corresponding scales based on number counts (colored regions, $\theta_{\ker}=$ 4, 8, and 16 arcmin) and the \MRLens\ case (solid and dashed contours). The Gaussian and count-based starlet cases yield almost identical constraints. Between four cases, the best result is given by the noise-based starlet case. Black stars represent the input cosmology. Gray zones are excluded in this analysis.}
	\label{fig:contour_comp_fixedAxis2_star1}
\end{figure*}

Next, we use all three Gaussian scales as the reference for the comparisons with the starlet function. As mentioned in \sect{sect:methodology_filters}, for the Gaussian filter scales of 1.2, 2.4, and 4.8 arcmin, we chose scales for the starlet based on two criteria: for an equal noise level, these are 2, 4, and 8 arcmin, and for equal number counts the corresponding scales are 4, 8, and 16 arcmin. The results are shown in \fig{fig:contour_comp_fixedAxis2_star1}. For the equal-number-count criterion, we see that if each scale gives approximately the same number of peaks, the $\OmegaM$-$\sigEig$ constraints obtained from the Gaussian and the starlet are similar (colored regions in the left and right panels). However, the starlet kernel leads to tighter constraints than the Gaussian when we match the same noise levels (lines and colored regions in the left panel). This results suggests that compensated kernels could be more powerful to extract cosmological information than non-compensated filters.

\begin{figure}[tb]
	\centering
	\includegraphics[width=\columnwidth]{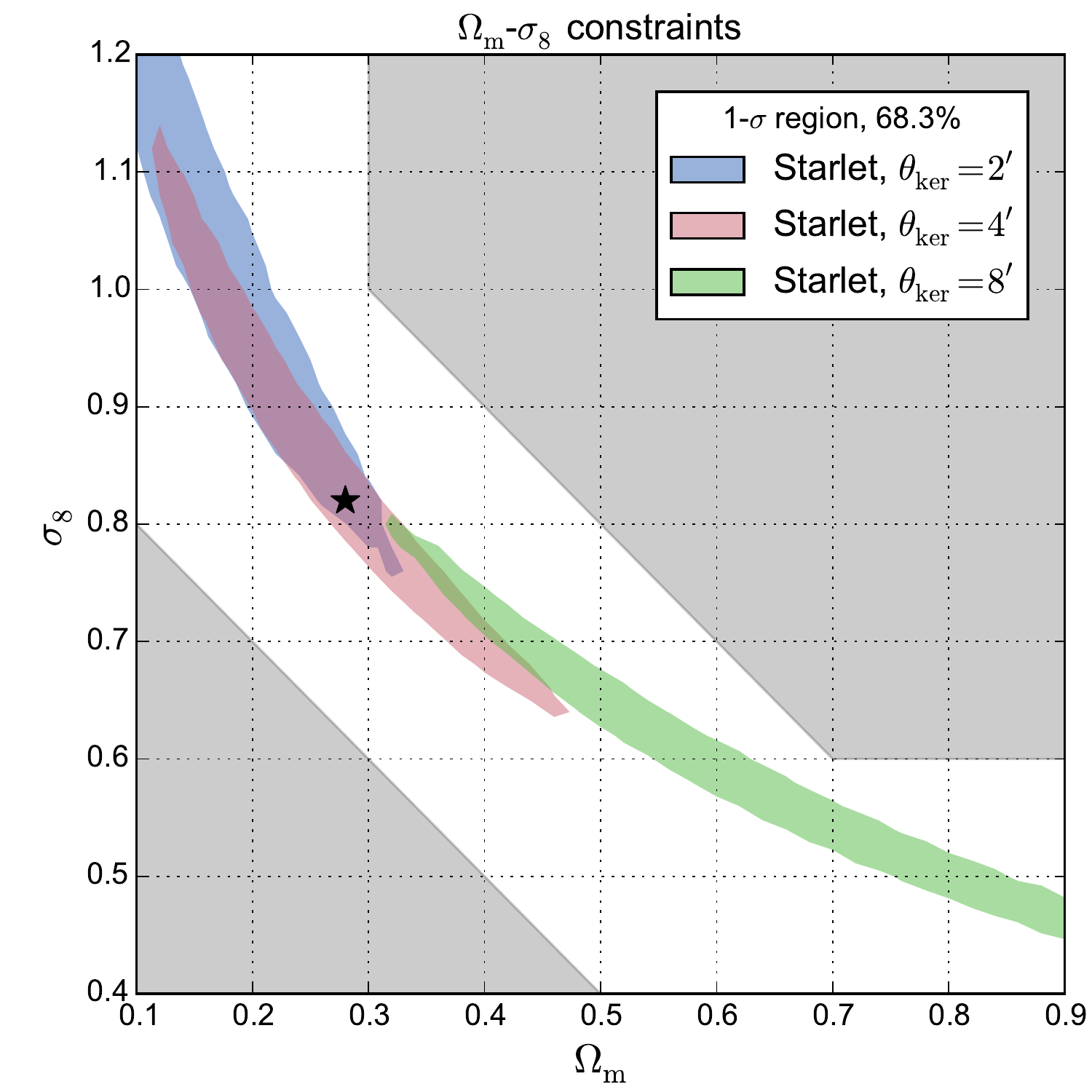}
	\caption{$\OmegaM$-$\sigEig$ 1-$\sigma$ region from individual scales of the starlet kernel. The plotted scales are 2 (blue), 4 (red), and 8 arcmin (green). The fact that different regions are preferred by different scales is more likely to be due to stochasticity. Gray zones are excluded in this analysis.}
	\label{fig:contour_comp_fixedAxis2_star3}
\end{figure}

We also draw constraints from individual scales of the starlet filter in (\fig{fig:contour_comp_fixedAxis2_star3}). It shows a very different behavior and seems to suggest that different scales could be sensitive to different cosmologies. However, this is actually a stochastic effect. We verify this statement by redoing the constraints with other observation vectors. It turns out that the scale-dependent tendency disappears. Nevertheless, when different cosmologies are prefered by different scales, the effect is less pronounced for the Gaussian filter. This is likely to be due to the fact that the starlet is a compensated filter, which is a band-pass function in the Fourier space. Since different filtering scales could be sensitive to different mass ranges of the mass function, band-pass filters could have a greater potential to separate better the multiscale information. The stochasticity of the observation vector suggests that the simulated field of view is rather small. While this should not affect the filter comparison nor contour sizes, actual cosmological constraints seem to require substantially larger data sets.

The right panel of \fig{fig:contour_comp_fixedAxis2_star1} shows the constraints from nonlinear filtering using \MRLens\ (solid and dashed lines). We observe that \MRLens\ conserves a strong degeneracy between $\OmegaM$ and $\sigEig$. The reasons for this result are various. First, large and small scales tend to sensitive to different halo masses which could help break the degeneracy. Using the combined strategy loses this advantage. Second, we have chosen a strict FDR. This rules out most of the spurious peaks, but also a lot of the signal. Third, as mentioned before, it is inappropriate to define signal-to-noise ratio when the filter is not linear. As a consequence, it is hardly possible to find bins for $\kappa$ peaks which are equivalent to $\nu$ bins in linear filtering. This is supported by \fig{fig:allFilters}, where we observe less peaks in the \MRLens\ map than in the other maps. Last, because of a low number of peaks, the binwidths need to be enlarged to contain larger number counts and to get closer to a Gaussian distribution, and large binwidths also weaken the signal.

A possible solution for exploring the \MRLens\ technique is to enhance the FDR and to redesign the binning. By increasing the number of peaks, thinner bins would be allowed. Another solution to better account for rare events in the current configuration is to use the Poisson likelihood. Finally, one could adopt the separated strategy, that is turning back to the methodology used by \citet{Pires_etal_2009a} that consists in estimating the peak abundance in the different scales before final reconstruction. Our comparison between linear and nonlinear techniques is basically the one between the separated and combined strategies.

\begin{table}
	\centering
	\caption{Quality indicators for $\OmegaM$-$\sigEig$ constraints with likelihood. All cases figured below use number counts on $g$ peaks. The quantity $\DCSE$ stands for the width of the contour, while the FoM is related to the area. In our study, combining five scales of starlet yield the best result in terms of FoM.}
	\begin{tabular}{clcc}
		\hline\hline\\[-2ex]
		Filter   & $\theta_{\ker}$ [arcmin] or $\alpha$  & $\DCSE$ & FoM\\
		\hline\\[-1.8ex]
		Gaussian & $\theta_{\ker}=$ 1.2                 & 0.045   & 19.1\\
		Gaussian & $\theta_{\ker}=$ 1.2, 2.4, 4.8       & 0.046   & 20.7\\
		Starlet  & $\theta_{\ker}=$ 2, 4, 8             & 0.046   & 23.4\\
		Starlet  & $\theta_{\ker}=$ 4, 8, 16            & 0.044   & 21.2\\
		Starlet  & $\theta_{\ker}=$ 2, 4, 8, 12, 16     & 0.045   & 24.8\\
		\MRLens  & $\alpha=$ 0.05                       & 0.046   & 16.2\\
		\hline
	\end{tabular}
	\label{tab:FoM_likelihood}
\end{table}

\tab{tab:FoM_likelihood} measures numerical qualities for constraints with different filtering techniques. It indicates that the width of contours does not vary significantly. The tightest constraint that we obtain is derived from a compensated filter.

\begin{figure}[tb]
	\centering
	\includegraphics[width=\columnwidth]{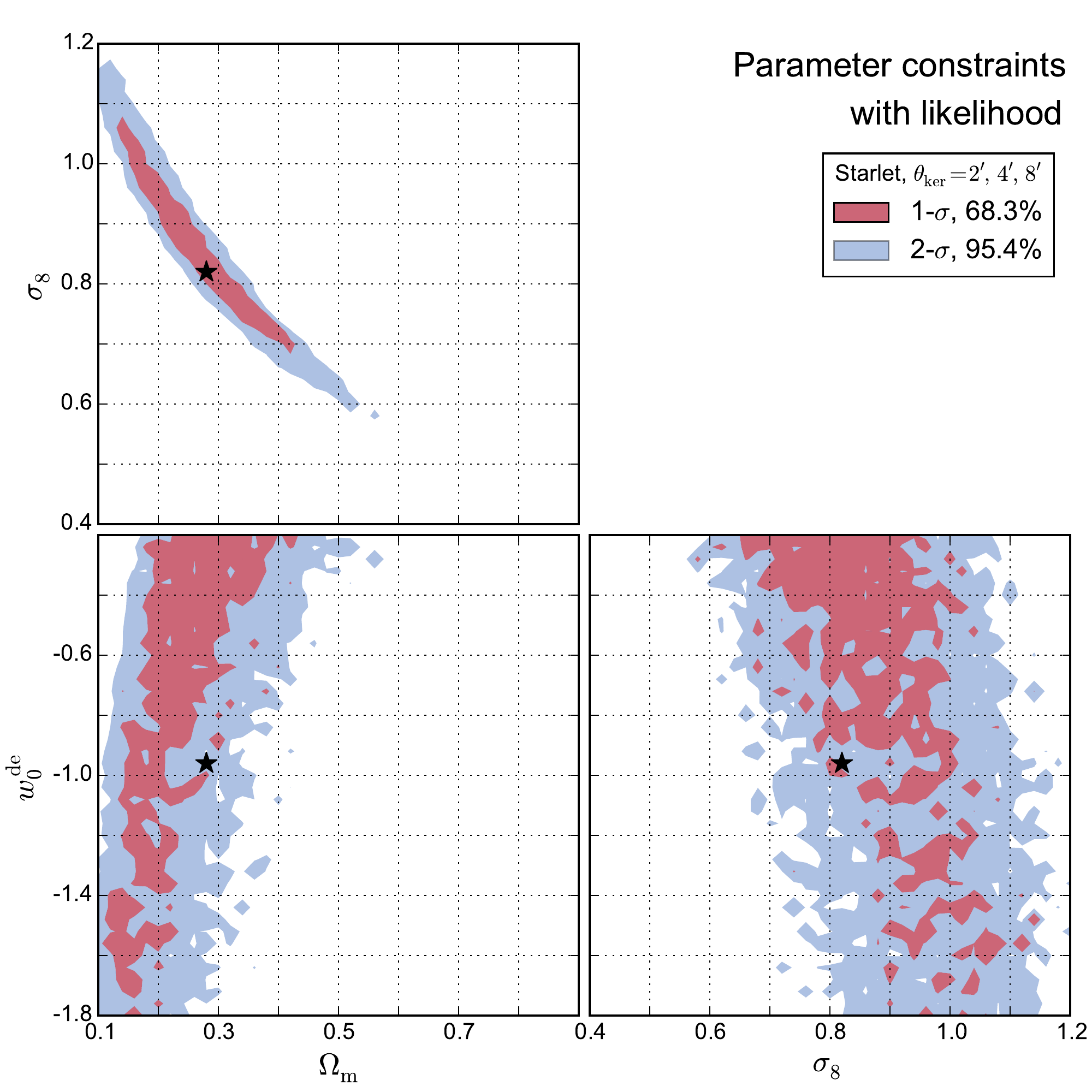}
	\caption{$\OmegaM$-$\sigEig$-$\wZero$ constraints using starlet with three scales. Each panel represents the contours derived from marginalized likelihood. Black stars are the input parameter values for the ``observation''. As far as $\wZero$ is concerned, the constraints are weak, but the degeneracies are clear. Fluctuations on both lower panels are due to usage of the copula likelihood.}
	\label{fig:contour_stair_star}
\end{figure}

Regarding results for $\wZero$, we show a representative case of starlet with $\theta_{\ker}=$ 2, 4, and 8 arcmin. \figFull{fig:contour_stair_star} presents the marginalized constraints of each doublet of parameters that we study. Those containing $\wZero$ are noisy because of the usage of the copula likelihood. We see that the current configuration of our model does not allow to impose constraints on $\wZero$. To measure this parameter, it could be useful to perform a tomography analysis to separate information of different stages of the late-time Universe. Nevertheless, our results successfully highlight the degeneracies of $\wZero$ with two other parameters. We fit the posterior density with:
\begin{align}
	I_1 &= \OmegaM - a_1 \wZero, \label{for:I_1}\\
	I_2 &= \sigEig + a_2 \wZero. \label{for:I_2}
\end{align}
We obtain for the slopes $a_1=0.108$ and $a_2=0.128$ for \fig{fig:contour_stair_star}. The results for the other filter functions are similar.

\subsection{Results from PMC ABC}
%\label{sect:results:ABC}

\begin{figure}[tb]
	\centering
	\includegraphics[width=\columnwidth]{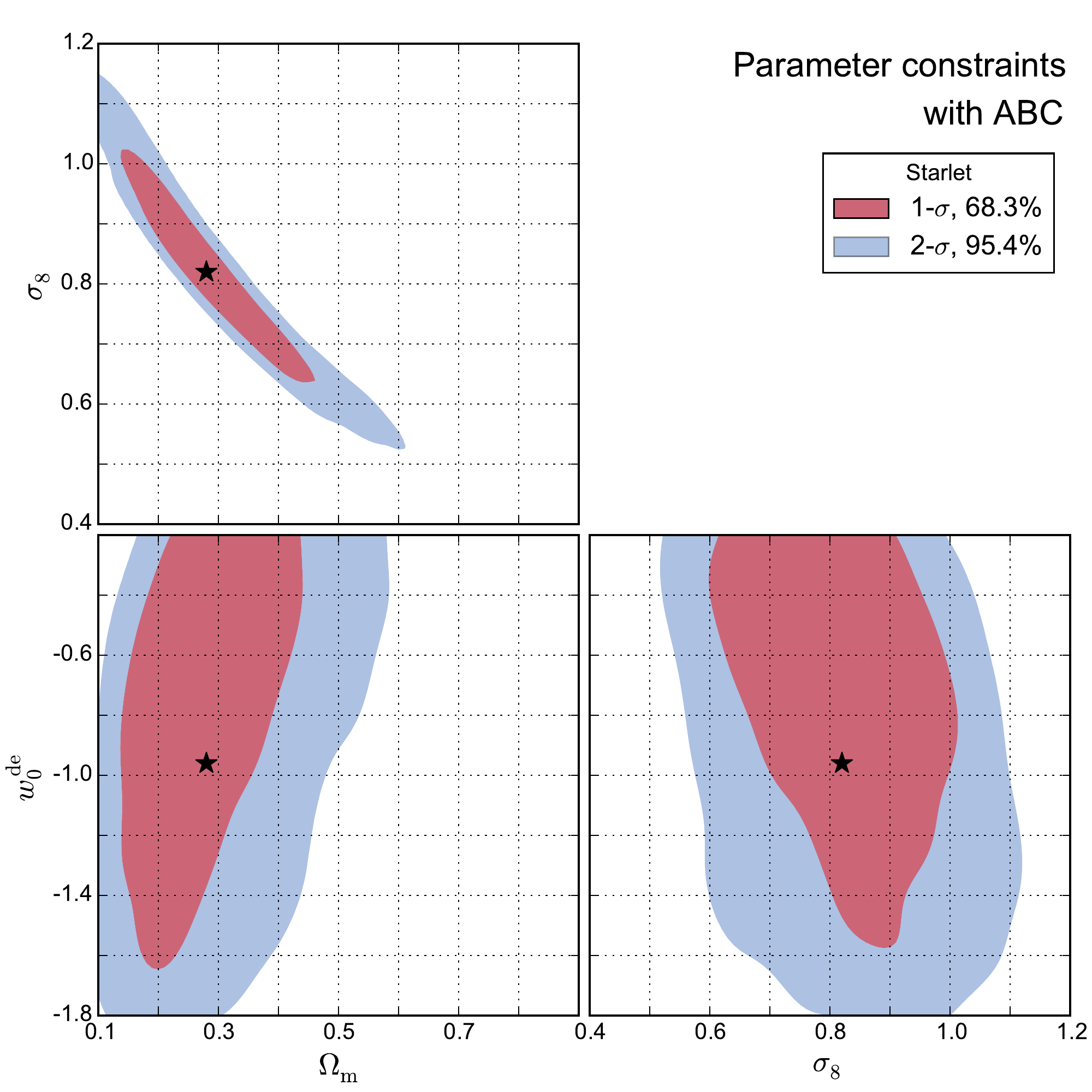}
	\caption{ABC constraints on $\OmegaM$, $\sigEig$, and $\wZero$ using starlet. The distance $D_2$ is used for this run. On each panel, the ABC posterior is marginalized over one of the three parameters. Black stars are the input cosmology.}
	\label{fig:contour_stair_ABC_star}
\end{figure}

We perform parameter constraints using the PMC ABC algorithm for our four cases. In \fig{fig:contour_stair_ABC_star}, we show the results derived from the starlet case using the fully correlated distance $D_2$. The contours are marginalized posteriors for all three pairs of parameters. They show the same degeneracy as we have found in \sect{sect:results:filtering}. We measure $a_1$ and $a_2$ from the ABC posteriors and obtain $a_1=0.083$ and $a_2=0.084$.

\begin{table}
	\centering
	\caption{Quality indicators for $\OmegaM$-$\sigEig$ constraints with PMC ABC. The quantity $\DCSE$ stands for the width of the contour, while the FoM is related to the area. ABC is used with two different distances $D_1$ and $D_2$ respectively given by Eqs. \eqref{for:D_1} and \eqref{for:D_2}. Here, we also put values from likelihood constraints using the same scales in this table for comparison. The kernel sizes for linear methods are defined in \tab{tab:x_mod_ABC}.}
	\begin{tabular}{cccc}
		\hline\hline\\[-2ex]
		Filter       & Constraints & $\DCSE$ & FoM\\
		\hline\\[-1.8ex]
		Gaussian     & Likelihood  & 0.046   & 20.7\\
		Gaussian     & ABC, $D_1$  & 0.043   & 16.3\\
		Gaussian     & ABC, $D_2$  & 0.059   & 11.7\\
		Starlet      & Likelihood  & 0.054   & 23.4\\
		Starlet      & ABC, $D_1$  & 0.050   & 15.5\\
		Starlet      & ABC, $D_2$  & 0.054   & 15.7\\
		$M_\ap$ tanh & ABC, $D_1$  & 0.037   & 19.4\\
		$M_\ap$ tanh & ABC, $D_2$  & 0.043   & 15.5\\
		\MRLens      & Likelihood  & 0.046   & 16.2\\
		\MRLens      & ABC, $D_1$  & 0.045   & 11.5\\
		\MRLens      & ABC, $D_2$  & 0.045   & 12.5\\
		\hline
	\end{tabular}
	\label{tab:FoM_ABC}
\end{table}

\begin{figure}[tb]
	\centering
	\includegraphics[width=\columnwidth]{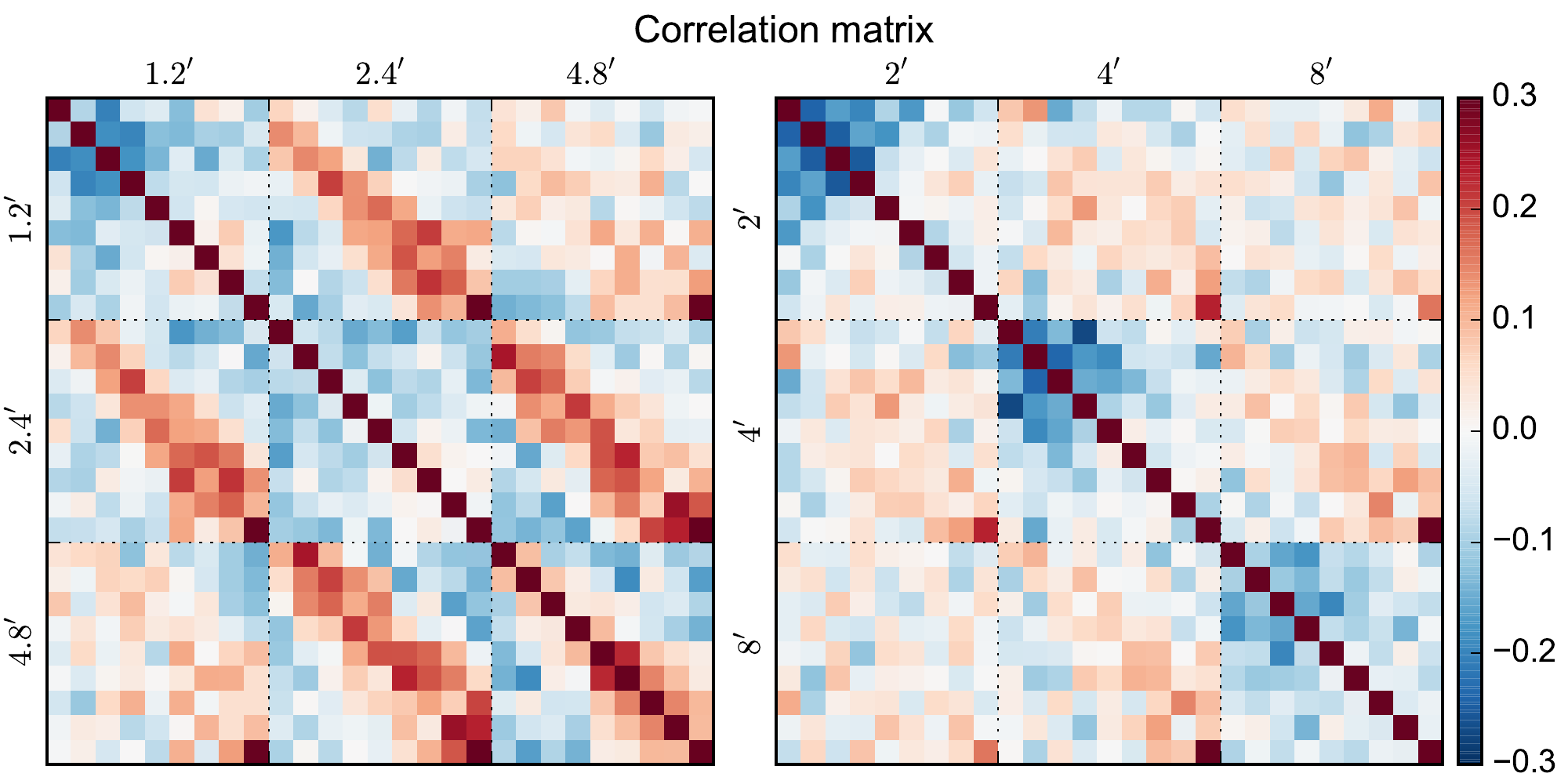}
	\caption{Correlation coefficient matrices under the input cosmology. \textit{Left panel}: the Gaussian case with $\theta_{\ker}=1.2$, 2.4, and 4.8 arcmin. \textit{Right panel}: the starlet case with $\theta_{\ker}=2$, 4, and 8 arcmin. Each of the 3$\times$3 blocks corresponds to the correlations between two filter scales. With each block, the S/N bins are [1, 1.5, 2, $\ldots$, 5, +$\infty$[. The data vector by starlet is less correlated.}
	\label{fig:corr_skip2}
\end{figure}

Using the same starlet filters, we compare two distances used for PMC ABC runs. When $D_1$ is used with the starlet, which means that data are treated as if uncorrelated, we find that the contour sizes do not change (see \tab{tab:FoM_ABC}) compared to $D_2$. For the Gaussian case, however, constraints from $D_1$ are tighter than those from $D_2$. This phenomenon is due to the off-diagonal elements of the covariance matrix. For non-compensated filters, the cross-correlations between bins are much stronger, as shown by \fig{fig:corr_skip2}. If these cross-correlations are ignored, the repeated peak counts in different bins are not properly accounted for. This overestimates the additional sensitivity to massive structures, and therefore produces overly tight constraints. As shown in \fig{fig:corr_skip2}, in the Gaussian case, adjacent filter scales show a 20--30\% correlation. The blurring of the off-diagonal stripes indicate a leakage to neighboring S/N bins due to noise, and the fact that clusters produce WL peaks with different S/N for different scales. On the contrary, in the case of the starlet, except for the highest S/N bin there are negligible correlations between different scales.

\begin{figure}[tb]
	\centering
	\includegraphics[width=\columnwidth]{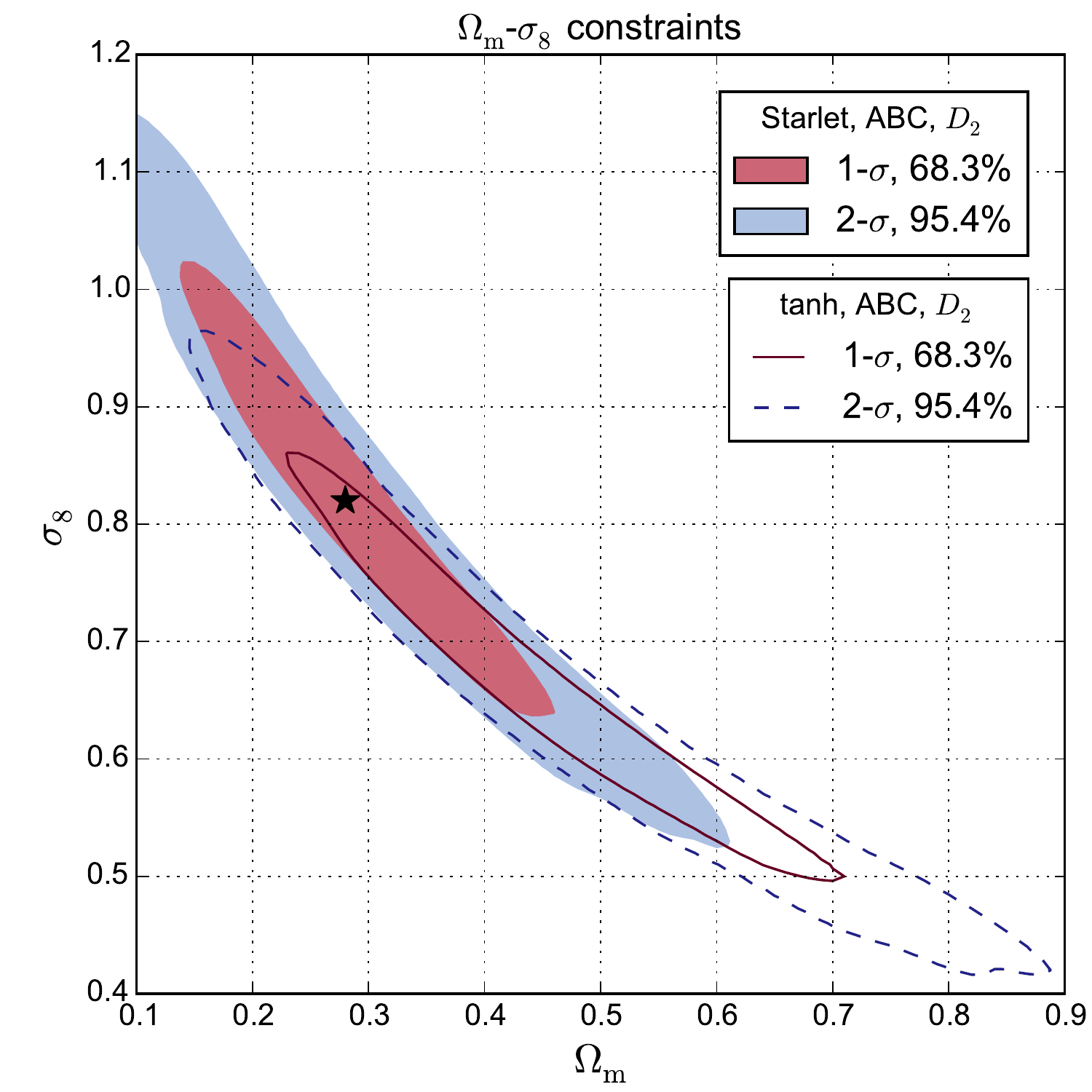}
	\caption{ABC $\OmegaM$-$\sigEig$ constraints from the starlet and the aperture mass. The distance $D_2$ is used in both cases. The black star is the input cosmology. The difference between two cases is that another observation data vector is created for the aperture mass and the direct comparison is not valid anymore.}
	\label{fig:contour_ABC_star_vs_tanh}
\end{figure}

\tab{tab:FoM_ABC} shows the ABC constraints from both the aperture mass and the starlet. We find that the FoM are close. However, in \fig{fig:contour_ABC_star_vs_tanh}, we see that the contours from the aperture mass is shifted toward high-$\OmegaM$ regions. The explanation for this shift is once again the stochasticity. We simulated another observation data vector for $M_\ap$, and the maximum-likelihood point for different methods do not coincide.

From \tab{tab:FoM_ABC}, one can see that the difference between \MRLens\ and linear filters using ABC is similar to using the likelihood. This suggests once again that the combined strategy leads to less tighter constraints than the separated strategy. Note that we also try to adjust $\alpha$ and run PMC ABC. However, without modifying the $\kappa$ bin choice, the resulting constraints do not differ substantially from $\alpha=0.05$.

\begin{figure*}[tb]
	\centering
	\includegraphics[width=8.5cm]{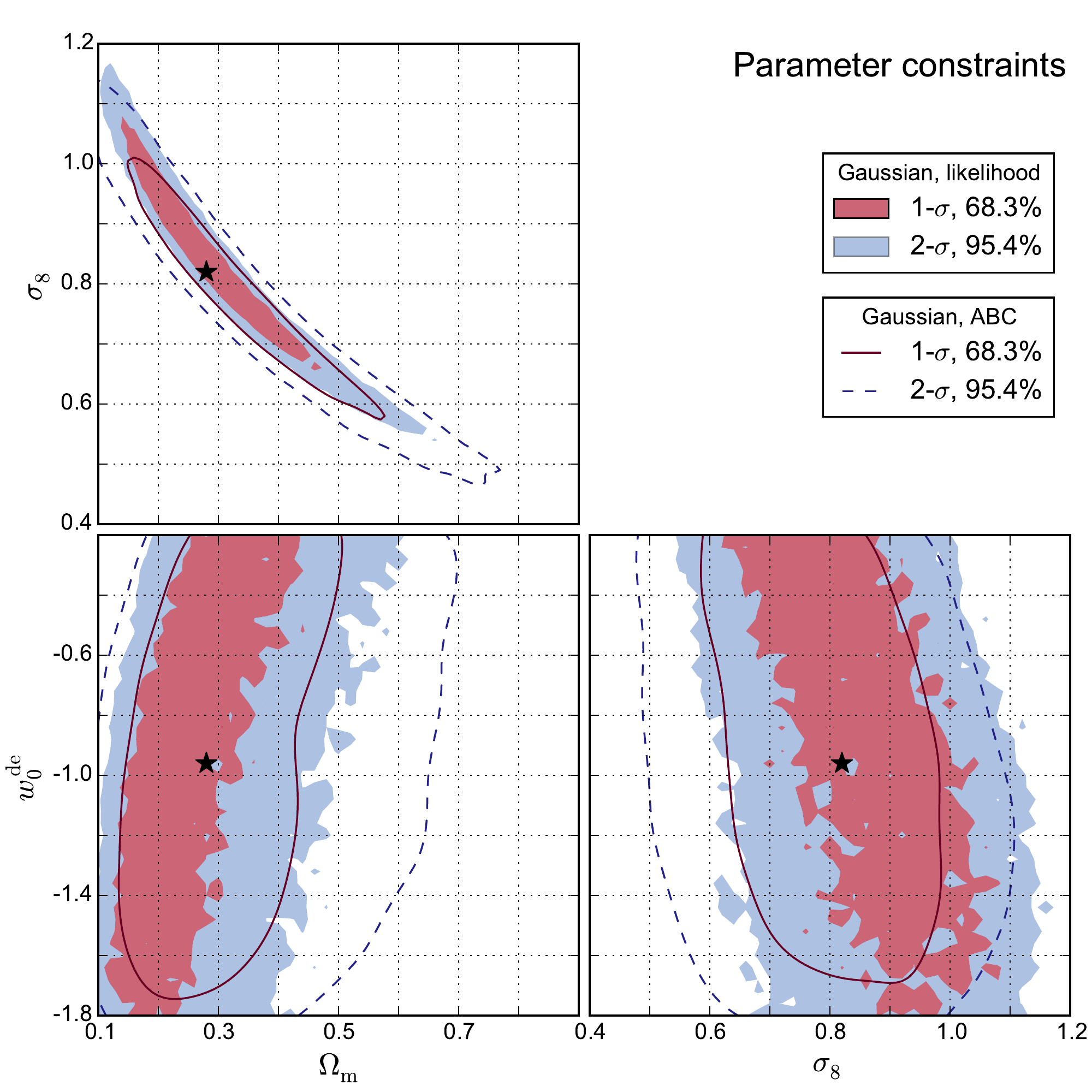}
	\includegraphics[width=8.5cm]{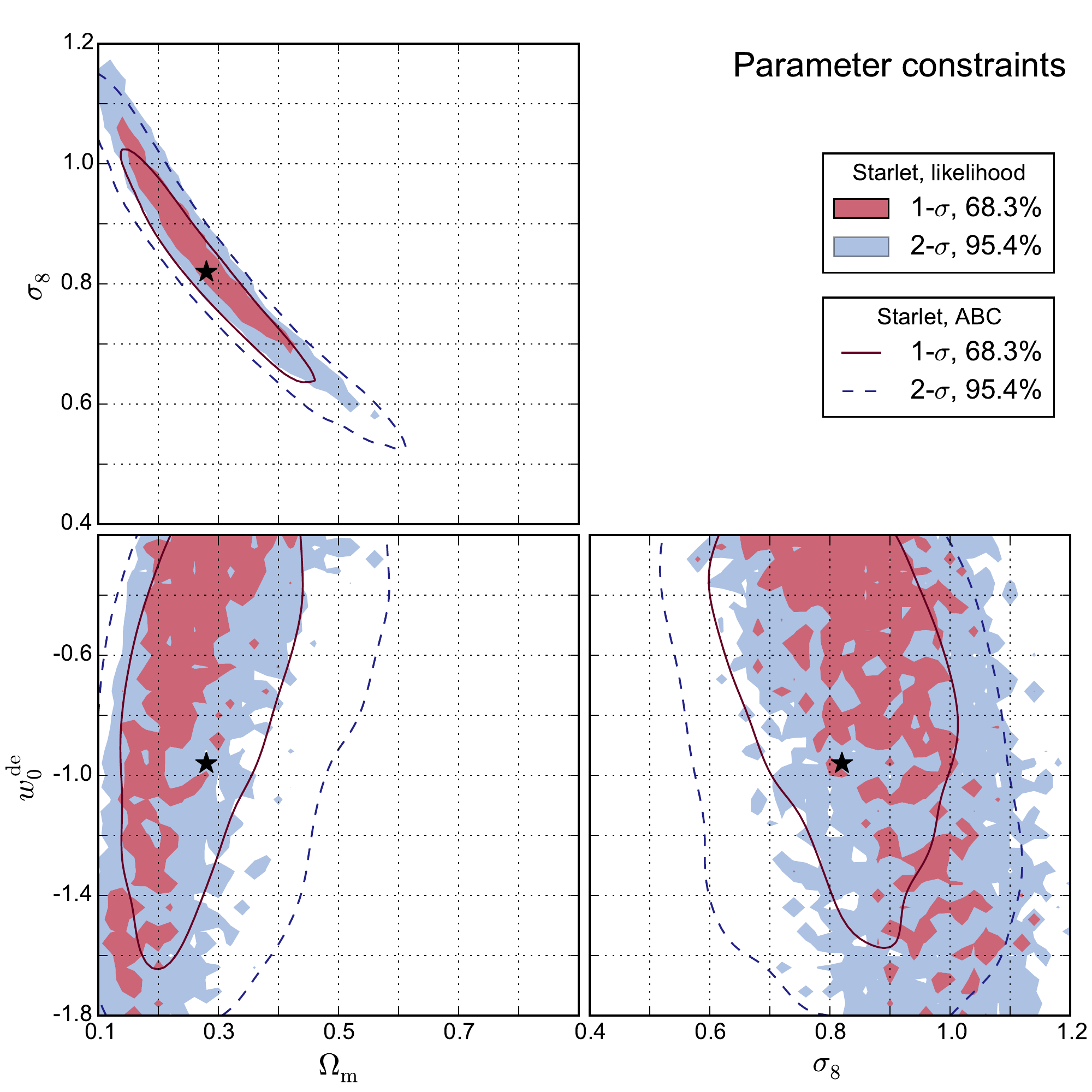}
	\caption{Comparison of $\OmegaM$-$\sigEig$-$\wZero$ constraints between likelihood and ABC. \textit{Left panel}: constraints with Gaussian smoothing. \textit{Right panel}: constraints with starlet filtering. Although ABC tolerates higher $\OmegaM$ and lower $\sigEig$ in both cases, two methods agree with each other.}
	\label{fig:contour_stair_comp_gauss}
\end{figure*}

Finally, we show the likelihood and ABC constraint contours for the Gaussian and starlet cases in \fig{fig:contour_stair_comp_gauss}. It turns out that ABC contours are systematically larger in the high-$\OmegaM$, low-$\sigEig$ region. This phenomenon was not observed in \PaperII\ where a similar comparison was made. We speculate that by including a third parameter $\wZero$ the contour becomes less precise, and ABC might be more sensitive to this effect. Note also that KDE is a biased estimator of posteriors (\PaperII). It smoothes the posterior and makes contours broader. Nevertheless, the ABC and likelihood constraints agree with each other. To be free from the bias, a possible alternative is to map the samples to a Gaussian distribution via some nonlinear mapping techniques \citep{Schuhmann_etal_2016}.

\section{Summary and discussion}
\label{sect:summary}

In this work, we studied WL peak counts for cosmology-oriented purposes. This means that we do not compare WL peaks with clusters of galaxies or study cluster properties, but use directly the peak abundance to constrain cosmological parameters.

We tested different filtering techniques by using our stochastic model for WL peak counts. The goal is not to find out the optimal filter, but to provide a standard procedure for comparison and to establish a roadmap for future studies. We claim that, rather than other indicators such as completeness and purity, it is more fair to study directly the tightness of constraint contours in order to maximize cosmological information extraction. In this work, we applied this principle with both the likelihood and ABC.

We compared Gaussian smoothing to starlet filtering, which is a comparison between compensated and non-compensated filters. Our results suggest that compensated filters are more suitable to capture cosmological information than the non-compensated ones. This comes from the fact that band-pass functions better separate multiscale information.

To handle multiscale data, we explored two strategies: the combined strategy creates a single mass map and the associated peak-count histogram; the separated strategy chooses some characteristic scales, produces one histogram per scale, and concatenates them into one data vector. We compared starlet filtering using the separated strategy with a nonlinear filter \MRLens, for which data were arranged using the combined strategy. The combined strategy, which mixes information of all scales yielded more elongated contours.

Concerning nonlinear methods, we would like to highlight that the linear-nonlinear comparison often contains a part of the separated-combined comparison. Although we did not carry out separate comparisons with regard to these two concepts in this work, some evidences still suggest that the separated-combined duality could be more influential than  the linear-nonlinear issue. A possible design for the separated-combined comparison is the comparison between the matched filter of \citet{Hennawi_Spergel_2005} and the multiscale aperture mass, or between \MRLens\ used as \citet{Pires_etal_2009a} with our \MRLens\ case. Also, the comparison between \MRLens\ used as \citet{Pires_etal_2009a} and linear starlet filtering can properly test the impact of nonlinear filters on constraints. However, this difference could be minor.

In this work, we found tighter constraints from the likelihood than from ABC. Since we had found in \PaperII\ that the copula likelihood closely approximates the true one, the constraints from ABC probably overestimate the true parameter errors. Using ABC, we performed parameter constraints for the aperture mass. This yielded a very similar FoM compared to the starlet, both compensated filters.

Concering the equation of state of dark energy, our results could not constrain $\wZero$ in general since $\wZero$ is degenerated with $\OmegaM$ and $\sigEig$. We fit these degeneracies with linear relations $I_1=\OmegaM-a_1\wZero$ and $I_2=\sigEig+a_2\wZero$ and found $a_1=0.108$ and $a_2=0.128$.

Our model for weak-lensing peak counts has been improved to be adapted to more realistic observational conditions. We have shown that our model is very general, and can be applied to weak-lensing data that is processed with conceptually different filtering approaches.

\begin{acknowledgements}
	This work is supported by Région d’Île-de-France with the DIM-ACAV thesis fellowship. We acknowledge the anonymous referee for reporting questions and comments. We also acknowledge support from the French national program for cosmology and galaxies (PNCG). The computing support from the in2p3 Computing Centre is thanked gratefully. Chieh-An Lin would like to thank Sarah Antier, C\'ecile Chenot, Samuel Farrens, Marie Gay, Zolt\'an Haiman, Olivier Iffrig, Ming Jiang, Fran\c{c}ois Lanusse, Yueh-Ning Lee, Sophia Lianou, Jia Liu, Fred Ngol\`e, Austin Peel, Jean-Luc Starck, and Jose Manuel Zorrilla Matilla for useful discussions, suggestions on diverse subjects, and technical support.
\end{acknowledgements}

\bibliography{Bibliographie_Linc}

\begin{thebibliography}{93}
\expandafter\ifx\csname natexlab\endcsname\relax\def\natexlab#1{#1}\fi

\bibitem[{{Abate} {et~al.}(2009){Abate}, {Wittman}, {Margoniner}, {Bridle},
  {Gee}, {Tyson}, \& {Dell'Antonio}}]{Abate_etal_2009}
{Abate}, A., {Wittman}, D., {Margoniner}, V.~E., {et~al.} 2009, \apj, 702, 603

\bibitem[{{Bard} {et~al.}(2013){Bard}, {Kratochvil}, {Chang}, {May}, {Kahn},
  {AlSayyad}, {Ahmad}, {Bankert}, {Connolly}, {Gibson}, {Gilmore}, {Grace},
  {Haiman}, {Hannel}, {Huffenberger}, {Jernigan}, {Jones}, {Krughoff},
  {Lorenz}, {Marshall}, {Meert}, {Nagarajan}, {Peng}, {Peterson}, {Rasmussen},
  {Shmakova}, {Sylvestre}, {Todd}, \& {Young}}]{Bard_etal_2013}
{Bard}, D., {Kratochvil}, J.~M., {Chang}, C., {et~al.} 2013, \apj, 774, 49

\bibitem[{{Bartelmann} {et~al.}(1996){Bartelmann}, {Narayan}, {Seitz}, \&
  {Schneider}}]{Bartelmann_etal_1996}
{Bartelmann}, M., {Narayan}, R., {Seitz}, S., \& {Schneider}, P. 1996, \apjl,
  464, L115

\bibitem[{{Bartelmann} {et~al.}(2002){Bartelmann}, {Perrotta}, \&
  {Baccigalupi}}]{Bartelmann_etal_2002}
{Bartelmann}, M., {Perrotta}, F., \& {Baccigalupi}, C. 2002, \aap, 396, 21

\bibitem[{{Baumann} {et~al.}(2012){Baumann}, {Nicolis}, {Senatore}, \&
  {Zaldarriaga}}]{Baumann_etal_2012}
{Baumann}, D., {Nicolis}, A., {Senatore}, L., \& {Zaldarriaga}, M. 2012, \jcap,
  7, 051

\bibitem[{Benjamini \& Hochberg(1995)}]{Benjamini_Hochberg_1995}
Benjamini, Y. \& Hochberg, Y. 1995, Journal of the Royal Statistical Society,
  Series B, 57, 289–300

\bibitem[{{Bernardeau} {et~al.}(2002){Bernardeau}, {Colombi}, {Gazta{\~n}aga},
  \& {Scoccimarro}}]{Bernardeau_etal_2002}
{Bernardeau}, F., {Colombi}, S., {Gazta{\~n}aga}, E., \& {Scoccimarro}, R.
  2002, \physrep, 367, 1

\bibitem[{Bhattacharya {et~al.}(2011)Bhattacharya, Heitmann, White, Luki{\'c},
  Wagner, \& Habib}]{Bhattacharya_etal_2011}
Bhattacharya, S., Heitmann, K., White, M., {et~al.} 2011, \apj, 732, 122

\bibitem[{{Bobin} {et~al.}(2014){Bobin}, {Sureau}, {Starck}, {Rassat}, \&
  {Paykari}}]{Bobin_etal_2014}
{Bobin}, J., {Sureau}, F., {Starck}, J.-L., {Rassat}, A., \& {Paykari}, P.
  2014, \aap, 563, A105

\bibitem[{Bourguignon {et~al.}(2011)Bourguignon, Mary, \&
  Slezak}]{Bourguignon_etal_2011}
Bourguignon, S., Mary, D., \& Slezak, E. 2011, Selected Topics in Signal
  Processing, IEEE Journal of, 5, 1002

\bibitem[{{Bullock} {et~al.}(2001){Bullock}, {Kolatt}, {Sigad}, {Somerville},
  {Kravtsov}, {Klypin}, {Primack}, \& {Dekel}}]{Bullock_etal_2001}
{Bullock}, J.~S., {Kolatt}, T.~S., {Sigad}, Y., {et~al.} 2001, \mnras, 321, 559

\bibitem[{{Cameron} \& {Pettitt}(2012)}]{Cameron_Pettitt_2012}
{Cameron}, E. \& {Pettitt}, A.~N. 2012, \mnras, 425, 44

\bibitem[{Candes \& Tao(2006)}]{Candes_Tao_2006}
Candes, E. \& Tao, T. 2006, Information Theory, IEEE Transactions on, 52, 5406

\bibitem[{Cand{\`e}s {et~al.}(2008)Cand{\`e}s, Wakin, \&
  Boyd}]{Candes_etal_2008}
Cand{\`e}s, E.~J., Wakin, M.~B., \& Boyd, S.~P. 2008, Journal of Fourier
  Analysis and Applications, 14, 877

\bibitem[{{Carrasco} {et~al.}(2012){Carrasco}, {Hertzberg}, \&
  {Senatore}}]{Carrasco_etal_2012}
{Carrasco}, J.~J.~M., {Hertzberg}, M.~P., \& {Senatore}, L. 2012, Journal of
  High Energy Physics, 9, 82

\bibitem[{{Carrillo} {et~al.}(2012){Carrillo}, {McEwen}, \&
  {Wiaux}}]{Carrillo_etal_2012}
{Carrillo}, R.~E., {McEwen}, J.~D., \& {Wiaux}, Y. 2012, \mnras, 426, 1223

\bibitem[{Daubechies {et~al.}(2004)Daubechies, Defrise, \&
  De~Mol}]{Daubechies_etal_2004}
Daubechies, I., Defrise, M., \& De~Mol, C. 2004, Communications on Pure and
  Applied Mathematics, 57, 1413

\bibitem[{Dietrich \& Hartlap(2010)}]{Dietrich_Hartlap_2010}
Dietrich, J.~P. \& Hartlap, J. 2010, \mnras, 402, 1049

\bibitem[{Dolag {et~al.}(2004)Dolag, Bartelmann, Perrotta, Baccigalupi,
  Moscardini, Meneghetti, \& Tormen}]{Dolag_etal_2004}
Dolag, K., Bartelmann, M., Perrotta, F., {et~al.} 2004, \aap, 416, 853

\bibitem[{{Efstathiou} {et~al.}(1991){Efstathiou}, {Bernstein}, {Tyson},
  {Katz}, \& {Guhathakurta}}]{Efstathiou_etal_1991}
{Efstathiou}, G., {Bernstein}, G., {Tyson}, J.~A., {Katz}, N., \&
  {Guhathakurta}, P. 1991, \apjl, 380, L47

\bibitem[{Elad \& Aharon(2006)}]{Elad_Aharon_2006}
Elad, M. \& Aharon, M. 2006, Image Processing, IEEE Transactions on, 15, 3736

\bibitem[{{Erben} {et~al.}(2013){Erben}, {Hildebrandt}, {Miller}, {van
  Waerbeke}, {Heymans}, {Hoekstra}, {Kitching}, {Mellier}, {Benjamin}, {Blake},
  {Bonnett}, {Cordes}, {Coupon}, {Fu}, {Gavazzi}, {Gillis}, {Grocutt}, {Gwyn},
  {Holhjem}, {Hudson}, {Kilbinger}, {Kuijken}, {Milkeraitis}, {Rowe},
  {Schrabback}, {Semboloni}, {Simon}, {Smit}, {Toader}, {Vafaei}, {van Uitert},
  \& {Velander}}]{Erben_etal_2013}
{Erben}, T., {Hildebrandt}, H., {Miller}, L., {et~al.} 2013, \mnras, 433, 2545

\bibitem[{Fadili {et~al.}(2009)Fadili, Starck, \& Murtagh}]{Fadili_etal_2009}
Fadili, M., Starck, J.-L., \& Murtagh, F. 2009, The Computer Journal, 52, 64

\bibitem[{Fan {et~al.}(2010)Fan, Shan, \& Liu}]{Fan_etal_2010}
Fan, Z., Shan, H., \& Liu, J. 2010, \apj, 719, 1408

\bibitem[{{Fu} {et~al.}(2014){Fu}, {Kilbinger}, {Erben}, {Heymans},
  {Hildebrandt}, {Hoekstra}, {Kitching}, {Mellier}, {Miller}, {Semboloni},
  {Simon}, {Van Waerbeke}, {Coupon}, {Harnois-D{\'e}raps}, {Hudson}, {Kuijken},
  {Rowe}, {Schrabback}, {Vafaei}, \& {Velander}}]{Fu_etal_2014}
{Fu}, L., {Kilbinger}, M., {Erben}, T., {et~al.} 2014, \mnras, 441, 2725

\bibitem[{{Gavazzi} \& {Soucail}(2007)}]{Gavazzi_Soucail_2007}
{Gavazzi}, R. \& {Soucail}, G. 2007, \aap, 462, 459

\bibitem[{Hamana {et~al.}(2012)Hamana, Oguri, Shirasaki, \&
  Sato}]{Hamana_etal_2012}
Hamana, T., Oguri, M., Shirasaki, M., \& Sato, M. 2012, \mnras, 425, 2287

\bibitem[{{Hamana} {et~al.}(2015){Hamana}, {Sakurai}, {Koike}, \&
  {Miller}}]{Hamana_etal_2015}
{Hamana}, T., {Sakurai}, J., {Koike}, M., \& {Miller}, L. 2015, \pasj, 67, 34

\bibitem[{Hamana {et~al.}(2004)Hamana, Takada, \& Yoshida}]{Hamana_etal_2004}
Hamana, T., Takada, M., \& Yoshida, N. 2004, \mnras, 350, 893

\bibitem[{{Hartlap} {et~al.}(2007){Hartlap}, {Simon}, \&
  {Schneider}}]{Hartlap_etal_2007}
{Hartlap}, J., {Simon}, P., \& {Schneider}, P. 2007, \aap, 464, 399

\bibitem[{Hennawi \& Spergel(2005)}]{Hennawi_Spergel_2005}
Hennawi, J.~F. \& Spergel, D.~N. 2005, \apj, 624, 59

\bibitem[{{Hetterscheidt} {et~al.}(2005){Hetterscheidt}, {Erben}, {Schneider},
  {Maoli}, {van Waerbeke}, \& {Mellier}}]{Hetterscheidt_etal_2005}
{Hetterscheidt}, M., {Erben}, T., {Schneider}, P., {et~al.} 2005, \aap, 442, 43

\bibitem[{Heymans {et~al.}(2012)Heymans, {Van Waerbeke}, Miller, Erben,
  Hildebrandt, Hoekstra, Kitching, Mellier, Simon, Bonnett, Coupon, Fu,
  {Harnois-D{\'e}raps}, Hudson, Kilbinger, Kuijken, Rowe, Schrabback,
  Semboloni, {van Uitert}, Vafaei, \& Velander}]{Heymans_etal_2012}
Heymans, C., {Van Waerbeke}, L., Miller, L., {et~al.} 2012, \mnras, 427, 146

\bibitem[{Jenkins {et~al.}(2001)Jenkins, Frenk, White, Colberg, Cole, Evrard,
  Couchman, \& Yoshida}]{Jenkins_etal_2001}
Jenkins, A., Frenk, C.~S., White, S.~D.~M., {et~al.} 2001, \mnras, 321, 372

\bibitem[{Kaiser \& Squires(1993)}]{Kaiser_Squires_1993}
Kaiser, N. \& Squires, G. 1993, \apj, 404, 441

\bibitem[{Kaiser {et~al.}(1994)Kaiser, Squires, Fahlman, \&
  Woods}]{Kaiser_etal_1994}
Kaiser, N., Squires, G., Fahlman, G., \& Woods, D. 1994, in Clusters of
  Galaxies, Vol.~1, Atlantica S{\'e}guier Fronti{\`e}res, 269

\bibitem[{Kilbinger {et~al.}(2013)Kilbinger, Fu, Heymans, Simpson, Benjamin,
  Erben, {Harnois-D{\'e}raps}, Hoekstra, Hildebrandt, Kitching, Mellier,
  Miller, {Van Waerbeke}, Benabed, Bonnett, Coupon, Hudson, Kuijken, Rowe,
  Schrabback, Semboloni, Vafaei, \& Velander}]{Kilbinger_etal_2013}
Kilbinger, M., Fu, L., Heymans, C., {et~al.} 2013, \mnras, 430, 2200

\bibitem[{{Killedar} {et~al.}(2015){Killedar}, {Borgani}, {Fabjan}, {Dolag},
  {Granato}, {Meneghetti}, {Planelles}, \&
  {Ragone-Figueroa}}]{Killedar_etal_2015}
{Killedar}, M., {Borgani}, S., {Fabjan}, D., {et~al.} 2015, ArXiv e-prints
  [\eprint[arXiv]{1507.05617}]

\bibitem[{{Kratochvil} {et~al.}(2010){Kratochvil}, {Haiman}, \&
  {May}}]{Kratochvil_etal_2010}
{Kratochvil}, J.~M., {Haiman}, Z., \& {May}, M. 2010, \prd, 81, 043519

\bibitem[{{Kuijken} {et~al.}(2015){Kuijken}, {Heymans}, {Hildebrandt},
  {Nakajima}, {Erben}, {de Jong}, {Viola}, {Choi}, {Hoekstra}, {Miller}, {van
  Uitert}, {Amon}, {Blake}, {Brouwer}, {Buddendiek}, {Conti}, {Eriksen},
  {Grado}, {Harnois-D{\'e}raps}, {Helmich}, {Herbonnet}, {Irisarri},
  {Kitching}, {Klaes}, {La Barbera}, {Napolitano}, {Radovich}, {Schneider},
  {Sif{\'o}n}, {Sikkema}, {Simon}, {Tudorica}, {Valentijn}, {Verdoes Kleijn},
  \& {van Waerbeke}}]{Kuijken_etal_2015}
{Kuijken}, K., {Heymans}, C., {Hildebrandt}, H., {et~al.} 2015, \mnras, 454,
  3500

\bibitem[{{Lambert} {et~al.}(2006){Lambert}, {Pires}, {Ballot},
  {Garc{\'{\i}}a}, {Starck}, \& {Turck-Chi{\`e}ze}}]{Lambert_etal_2006}
{Lambert}, P., {Pires}, S., {Ballot}, J., {et~al.} 2006, \aap, 454, 1021

\bibitem[{{Lanusse} {et~al.}(2016){Lanusse}, {Starck}, {Leonard}, \&
  {Pires}}]{Lanusse_etal_2016}
{Lanusse}, F., {Starck}, J.-L., {Leonard}, A., \& {Pires}, S. 2016, ArXiv
  e-prints [\eprint[arXiv]{1603.01599}]

\bibitem[{{Leonard} {et~al.}(2014){Leonard}, {Lanusse}, \&
  {Starck}}]{Leonard_etal_2014}
{Leonard}, A., {Lanusse}, F., \& {Starck}, J.-L. 2014, \mnras, 440, 1281

\bibitem[{{Lin} \& {Kilbinger}(2015{\natexlab{a}})}]{Lin_Kilbinger_2015}
{Lin}, C.-A. \& {Kilbinger}, M. 2015{\natexlab{a}}, \aap, 576, A24 (Paper I)

\bibitem[{{Lin} \& {Kilbinger}(2015{\natexlab{b}})}]{Lin_Kilbinger_2015a}
{Lin}, C.-A. \& {Kilbinger}, M. 2015{\natexlab{b}}, \aap, 583, A70 (Paper II)

\bibitem[{{Liu} {et~al.}(2014){Liu}, {Haiman}, {Hui}, {Kratochvil}, \&
  {May}}]{Liu_etal_2014a}
{Liu}, J., {Haiman}, Z., {Hui}, L., {Kratochvil}, J.~M., \& {May}, M. 2014,
  \prd, 89, 023515

\bibitem[{{Liu} {et~al.}(2015{\natexlab{a}}){Liu}, {Petri}, {Haiman}, {Hui},
  {Kratochvil}, \& {May}}]{Liu_etal_2015}
{Liu}, J., {Petri}, A., {Haiman}, Z., {et~al.} 2015{\natexlab{a}}, \prd, 91,
  063507

\bibitem[{{Liu} {et~al.}(2015{\natexlab{b}}){Liu}, {Pan}, {Li}, {Shan}, {Wang},
  {Fu}, {Fan}, {Kneib}, {Leauthaud}, {Van Waerbeke}, {Makler}, {Moraes},
  {Erben}, \& {Charbonnier}}]{Liu_etal_2015a}
{Liu}, X., {Pan}, C., {Li}, R., {et~al.} 2015{\natexlab{b}}, \mnras, 450, 2888

\bibitem[{Liu {et~al.}(2014)Liu, Wang, Pan, \& Fan}]{Liu_etal_2014}
Liu, X., Wang, Q., Pan, C., \& Fan, Z. 2014, \apj, 784, 31

\bibitem[{{Makino} {et~al.}(1992){Makino}, {Sasaki}, \&
  {Suto}}]{Makino_etal_1992}
{Makino}, N., {Sasaki}, M., \& {Suto}, Y. 1992, \prd, 46, 585

\bibitem[{{Marian} {et~al.}(2011){Marian}, {Hilbert}, {Smith}, {Schneider}, \&
  {Desjacques}}]{Marian_etal_2011}
{Marian}, L., {Hilbert}, S., {Smith}, R.~E., {Schneider}, P., \& {Desjacques},
  V. 2011, \apjl, 728, L13

\bibitem[{{Marian} {et~al.}(2009){Marian}, {Smith}, \&
  {Bernstein}}]{Marian_etal_2009}
{Marian}, L., {Smith}, R.~E., \& {Bernstein}, G.~M. 2009, \apjl, 698, L33

\bibitem[{{Marian} {et~al.}(2010){Marian}, {Smith}, \&
  {Bernstein}}]{Marian_etal_2010}
{Marian}, L., {Smith}, R.~E., \& {Bernstein}, G.~M. 2010, \apj, 709, 286

\bibitem[{Marian {et~al.}(2012)Marian, Smith, Hilbert, \&
  Schneider}]{Marian_etal_2012}
Marian, L., Smith, R.~E., Hilbert, S., \& Schneider, P. 2012, \mnras, 423, 1711

\bibitem[{Marian {et~al.}(2013)Marian, Smith, Hilbert, \&
  Schneider}]{Marian_etal_2013}
Marian, L., Smith, R.~E., Hilbert, S., \& Schneider, P. 2013, \mnras, 432, 1338

\bibitem[{{Martinet} {et~al.}(2015){Martinet}, {Bartlett}, {Kiessling}, \&
  {Sartoris}}]{Martinet_etal_2015}
{Martinet}, N., {Bartlett}, J.~G., {Kiessling}, A., \& {Sartoris}, B. 2015,
  \aap, 581, A101

\bibitem[{Maturi {et~al.}(2010)Maturi, Angrick, Pace, \&
  Bartelmann}]{Maturi_etal_2010}
Maturi, M., Angrick, C., Pace, F., \& Bartelmann, M. 2010, \aap, 519, A23

\bibitem[{Maturi {et~al.}(2011)Maturi, Fedeli, \&
  Moscardini}]{Maturi_etal_2011}
Maturi, M., Fedeli, C., \& Moscardini, L. 2011, \mnras, 416, 2527

\bibitem[{{Maturi} {et~al.}(2005){Maturi}, {Meneghetti}, {Bartelmann}, {Dolag},
  \& {Moscardini}}]{Maturi_etal_2005}
{Maturi}, M., {Meneghetti}, M., {Bartelmann}, M., {Dolag}, K., \& {Moscardini},
  L. 2005, \aap, 442, 851

\bibitem[{Navarro {et~al.}(1996)Navarro, Frenk, \& White}]{Navarro_etal_1996}
Navarro, J.~F., Frenk, C.~S., \& White, S. D.~M. 1996, \apj, 462, 563

\bibitem[{Navarro {et~al.}(1997)Navarro, Frenk, \& White}]{Navarro_etal_1997}
Navarro, J.~F., Frenk, C.~S., \& White, S. D.~M. 1997, \apj, 490, 493

\bibitem[{{Ngol{\`e} Mboula} {et~al.}(2015){Ngol{\`e} Mboula}, {Starck},
  {Ronayette}, {Okumura}, \& {Amiaux}}]{NgoleMboula_etal_2015}
{Ngol{\`e} Mboula}, F.~M., {Starck}, J.-L., {Ronayette}, S., {Okumura}, K., \&
  {Amiaux}, J. 2015, \aap, 575, A86

\bibitem[{Pires {et~al.}(2012)Pires, Leonard, \& Starck}]{Pires_etal_2012}
Pires, S., Leonard, A., \& Starck, J.-L. 2012, \mnras, 423, 983

\bibitem[{{Pires} {et~al.}(2009{\natexlab{a}}){Pires}, {Starck}, {Amara},
  {R{\'e}fr{\'e}gier}, \& {Teyssier}}]{Pires_etal_2009a}
{Pires}, S., {Starck}, J.-L., {Amara}, A., {R{\'e}fr{\'e}gier}, A., \&
  {Teyssier}, R. 2009{\natexlab{a}}, \aap, 505, 969

\bibitem[{{Pires} {et~al.}(2009{\natexlab{b}}){Pires}, {Starck}, {Amara},
  {Teyssier}, {R{\'e}fr{\'e}gier}, \& {Fadili}}]{Pires_etal_2009}
{Pires}, S., {Starck}, J.-L., {Amara}, A., {et~al.} 2009{\natexlab{b}}, \mnras,
  395, 1265

\bibitem[{Press \& Schechter(1974)}]{Press_Schechter_1974}
Press, W.~H. \& Schechter, P. 1974, \apj, 187, 425

\bibitem[{{Robin} {et~al.}(2014){Robin}, {Reyl{\'e}}, {Fliri}, {Czekaj},
  {Robert}, \& {Martins}}]{Robin_etal_2014}
{Robin}, A.~C., {Reyl{\'e}}, C., {Fliri}, J., {et~al.} 2014, \aap, 569, A13

\bibitem[{Schirmer {et~al.}(2007)Schirmer, Erben, Hetterscheidt, \&
  Schneider}]{Schirmer_etal_2007}
Schirmer, M., Erben, T., Hetterscheidt, M., \& Schneider, P. 2007, \aap, 462,
  875

\bibitem[{{Schirmer} {et~al.}(2004){Schirmer}, {Erben}, {Schneider}, {Wolf}, \&
  {Meisenheimer}}]{Schirmer_etal_2004}
{Schirmer}, M., {Erben}, T., {Schneider}, P., {Wolf}, C., \& {Meisenheimer}, K.
  2004, \aap, 420, 75

\bibitem[{Schneider(1996)}]{Schneider_1996}
Schneider, P. 1996, \mnras, 283, 837

\bibitem[{Schneider {et~al.}(1998)Schneider, {Van Waerbeke}, Jain, \&
  Kruse}]{Schneider_etal_1998}
Schneider, P., {Van Waerbeke}, L., Jain, B., \& Kruse, G. 1998, \mnras, 296,
  873

\bibitem[{{Schuhmann} {et~al.}(2016){Schuhmann}, {Joachimi}, \&
  {Peiris}}]{Schuhmann_etal_2016}
{Schuhmann}, R.~L., {Joachimi}, B., \& {Peiris}, H.~V. 2016, \mnras, 459, 1916

\bibitem[{Seitz \& Schneider(1995)}]{Seitz_Schneider_1995}
Seitz, C. \& Schneider, P. 1995, \aap, 297, 287

\bibitem[{{Sellentin} \& {Heavens}(2016)}]{Sellentin_Heavens_2016}
{Sellentin}, E. \& {Heavens}, A.~F. 2016, \mnras, 456, L132

\bibitem[{Sheth \& Tormen(1999)}]{Sheth_Tormen_1999}
Sheth, R.~K. \& Tormen, G. 1999, \mnras, 308, 119

\bibitem[{Sheth \& Tormen(2002)}]{Sheth_Tormen_2002}
Sheth, R.~K. \& Tormen, G. 2002, \mnras, 329, 61

\bibitem[{Sklar(1959)}]{Sklar_1959}
Sklar, A. 1959, Publ. Inst. Statist. Univ. Paris, 8, 229

\bibitem[{Starck {et~al.}(2002)Starck, Murtagh, \& Fadili}]{Starck_etal_2002}
Starck, J.-L., Murtagh, F., \& Fadili, J.~M. 2002, Sparse Image and Signal
  Processing (Cambridge University Press)

\bibitem[{{Starck} {et~al.}(2006){Starck}, {Pires}, \&
  {R{\'e}fr{\'e}gier}}]{Starck_etal_2006}
{Starck}, J.-L., {Pires}, S., \& {R{\'e}fr{\'e}gier}, A. 2006, \aap, 451, 1139

\bibitem[{Takada \& Jain(2002)}]{Takada_Jain_2002}
Takada, M. \& Jain, B. 2002, \mnras, 337, 875

\bibitem[{Takada \& Jain(2003{\natexlab{a}})}]{Takada_Jain_2003a}
Takada, M. \& Jain, B. 2003{\natexlab{a}}, \mnras, 340, 580

\bibitem[{Takada \& Jain(2003{\natexlab{b}})}]{Takada_Jain_2003b}
Takada, M. \& Jain, B. 2003{\natexlab{b}}, \mnras, 344, 857

\bibitem[{{The Dark Energy Survey Collaboration} {et~al.}(2015){The Dark Energy
  Survey Collaboration}, {Abbott}, {Abdalla}, {Allam}, {Amara}, {Annis},
  {Armstrong}, {Bacon}, {Banerji}, {Bauer}, {Baxter}, {Becker},
  {Benoit-L{\'e}vy}, {Bernstein}, {Bernstein}, {Bertin}, {Blazek}, {Bonnett},
  {Bridle}, {Brooks}, {Bruderer}, {Buckley-Geer}, {Burke}, {Busha}, {Capozzi},
  {Carnero Rosell}, {Carrasco Kind}, {Carretero}, {Castander}, {Chang},
  {Clampitt}, {Crocce}, {Cunha}, {D'Andrea}, {da Costa}, {Das}, {DePoy},
  {Desai}, {Diehl}, {Dietrich}, {Dodelson}, {Doel}, {Drlica-Wagner},
  {Efstathiou}, {Eifler}, {Erickson}, {Estrada}, {Evrard}, {Fausti Neto},
  {Fernandez}, {Finley}, {Flaugher}, {Fosalba}, {Friedrich}, {Frieman},
  {Gangkofner}, {Garcia-Bellido}, {Gaztanaga}, {Gerdes}, {Gruen}, {Gruendl},
  {Gutierrez}, {Hartley}, {Hirsch}, {Honscheid}, {Huff}, {Jain}, {James},
  {Jarvis}, {Kacprzak}, {Kent}, {Kirk}, {Krause}, {Kravtsov}, {Kuehn},
  {Kuropatkin}, {Kwan}, {Lahav}, {Leistedt}, {Li}, {Lima}, {Lin}, {MacCrann},
  {March}, {Marshall}, {Martini}, {McMahon}, {Melchior}, {Miller}, {Miquel},
  {Mohr}, {Neilsen}, {Nichol}, {Nicola}, {Nord}, {Ogando}, {Palmese}, {Peiris},
  {Plazas}, {Refregier}, {Roe}, {Romer}, {Roodman}, {Rowe}, {Rykoff}, {Sabiu},
  {Sadeh}, {Sako}, {Samuroff}, {S{\'a}nchez}, {Sanchez}, {Seo},
  {Sevilla-Noarbe}, {Sheldon}, {Smith}, {Soares-Santos}, {Sobreira}, {Suchyta},
  {Swanson}, {Tarle}, {Thaler}, {Thomas}, {Troxel}, {Vikram}, {Walker},
  {Wechsler}, {Weller}, {Zhang}, \&
  {Zuntz}}]{TheDarkEnergySurveyCollaboration_etal_2015}
{The Dark Energy Survey Collaboration}, {Abbott}, T., {Abdalla}, F.~B.,
  {et~al.} 2015, ArXiv e-prints [\eprint[arXiv]{1507.05552}]

\bibitem[{{Tinker} {et~al.}(2008){Tinker}, {Kravtsov}, {Klypin}, {Abazajian},
  {Warren}, {Yepes}, {Gottl{\"o}ber}, \& {Holz}}]{Tinker_etal_2008a}
{Tinker}, J., {Kravtsov}, A.~V., {Klypin}, A., {et~al.} 2008, \apj, 688, 709

\bibitem[{{Van Waerbeke}(2000)}]{VanWaerbeke_2000}
{Van Waerbeke}, L. 2000, \mnras, 313, 524

\bibitem[{{Van Waerbeke} {et~al.}(2013){Van Waerbeke}, {Benjamin}, {Erben},
  {Heymans}, {Hildebrandt}, {Hoekstra}, {Kitching}, {Mellier}, {Miller},
  {Coupon}, {Harnois-D{\'e}raps}, {Fu}, {Hudson}, {Kilbinger}, {Kuijken},
  {Rowe}, {Schrabback}, {Semboloni}, {Vafaei}, {van Uitert}, \&
  {Velander}}]{VanWaerbeke_etal_2013}
{Van Waerbeke}, L., {Benjamin}, J., {Erben}, T., {et~al.} 2013, \mnras, 433,
  3373

\bibitem[{Wang {et~al.}(2009)Wang, Haiman, \& May}]{Wang_etal_2009}
Wang, S., Haiman, Z., \& May, M. 2009, \apj, 691, 547

\bibitem[{{Warren} {et~al.}(2006){Warren}, {Abazajian}, {Holz}, \&
  {Teodoro}}]{Warren_etal_2006}
{Warren}, M.~S., {Abazajian}, K., {Holz}, D.~E., \& {Teodoro}, L. 2006, \apj,
  646, 881

\bibitem[{{Weinberg} \& {Kamionkowski}(2003)}]{Weinberg_Kamionkowski_2003}
{Weinberg}, N.~N. \& {Kamionkowski}, M. 2003, \mnras, 341, 251

\bibitem[{{Weyant} {et~al.}(2013){Weyant}, {Schafer}, \&
  {Wood-Vasey}}]{Weyant_etal_2013}
{Weyant}, A., {Schafer}, C., \& {Wood-Vasey}, W.~M. 2013, \apj, 764, 116

\bibitem[{{White} {et~al.}(2002){White}, {van Waerbeke}, \&
  {Mackey}}]{White_etal_2002}
{White}, M., {van Waerbeke}, L., \& {Mackey}, J. 2002, \apj, 575, 640

\bibitem[{{Yang} {et~al.}(2013){Yang}, {Kratochvil}, {Huffenberger}, {Haiman},
  \& {May}}]{Yang_etal_2013}
{Yang}, X., {Kratochvil}, J.~M., {Huffenberger}, K., {Haiman}, Z., \& {May}, M.
  2013, \prd, 87, 023511

\bibitem[{Yang {et~al.}(2011)Yang, Kratochvil, Wang, Lim, Haiman, \&
  May}]{Yang_etal_2011}
Yang, X., Kratochvil, J.~M., Wang, S., {et~al.} 2011, \prd, 84, 043529

\end{thebibliography}

\end{document}